\documentstyle[prd,aps,preprint,tighten,epsfig]{revtex}
\begin{document}
\preprint{                                                 IASSNS-AST 96/41} 
\draft%......................................................................
\title{\Large\bf    Three-flavor atmospheric neutrino anomaly}
%............................................................................
%
\author{		G.~L.~Fogli$\,^b$,\ \  
			E.~Lisi$\,^{a,b}$,\ \ 
			D.~Montanino$\,^b$,\ \ and\ \ 
			G.~Scioscia$\,^b$}
\address{$^a$Institute for Advanced Study, Princeton, New Jersey 08540\\ 
         $^b$Dipartimento di Fisica dell'Universit{\`a}
	   and Sezione INFN di Bari, 70126 Bari, Italy}
\maketitle
\begin{abstract}
%............................................................................
We investigate the indications of flavor oscillations that come from the 
anomalous flavor composition of the atmospheric neutrino flux observed in 
some underground experiments. We study the information coming from the 
neutrino-induced $\mu$-like and $e$-like events both in the sub-GeV energy 
range (Kamiokande, IMB, Fr{\'e}jus, and NUSEX experiments) and in the 
multi-GeV energy range (Kamiokande experiment). First we analyze all the 
data in the limits of pure $\nu_\mu\leftrightarrow\nu_\tau$ and 
$\nu_\mu\leftrightarrow\nu_e$ oscillations. We obtain that 
$\nu_\mu\leftrightarrow\nu_e$ oscillations provide a better fit, in 
particular to the multi-GeV data. Then we perform a three-flavor analysis 
in the hypothesis of dominance of one neutrino square mass difference, 
$m^2$, implying that the neutrino mixing is parametrized by two angles, 
$(\psi,\,\phi)\in[0,\,\pi/2]$. We explore the space $(m^2,\,\psi,\,\phi)$ 
exhaustively, and find the regions favored by the oscillation hypothesis. 
The results are displayed in a form suited to the comparison with other 
flavor oscillation searches at accelerator, reactor, and solar $\nu$ 
experiments. In the analysis, we pay particular attention to the earth 
matter effects, to the correlation of the uncertainties, and to the 
symmetry properties of the oscillation probability.
%...........................................................................
\end{abstract}
%...........................................................................
%	14.60.Pq	Neutrino mass and mixing
%	95.85.Ry	Neutrino, muon, pion and other elem. particles
%	13.15.+g	Neutrino interactions
%%%%%%%%%%%%%%%%%%%%%%%%%%%%%%%%%%%%%%%%%%%%%%%%%%%%%%%%%%%%%%%%%%%%%%%%%%%% 
\section{INTRODUCTION}
%%%%%%%%%%%%%%%%%%%%%%%%%%%%%%%%%%%%%%%%%%%%%%%%%%%%%%%%%%%%%%%%%%%%%%%%%%%%

   	The indication for an anomalous muon and electron flavor 
composition of the observed atmospheric neutrino flux represents a still 
unsolved puzzle (for recent reviews, see \cite{Revw}). A possible 
explanation could be provided by neutrino flavor oscillations. In this 
paper we adopt such a viewpoint and explore systematically its 
consequences  in a three-flavor framework with one dominant neutrino 
square mass difference.

	We perform a comprehensive and accurate analysis of the 
experimental information coming from the Kamiokande \cite{Hi88,Hi92,Fu94}, 
IMB \cite{Ca91,Be92}, Frejus \cite{Be89,Be90}, and NUSEX \cite{Ag89} 
atmospheric neutrino experiments. We do not characterize the atmospheric 
neutrino anomaly with the popular double flavor ratio
$R_{\mu/e}=(\mu/e)_{\rm data}/(\mu/e)_{\rm theory}$ which, as discussed 
in \cite{Li95},  is affected by non-Gaussian uncertainties. Instead we 
separate the $\mu$-like and $e$-like event rates, whose errors are normally 
distributed \cite{Li95}. We compare the data with the theoretical 
expectations in the presence of two-flavor and three-flavor oscillations, 
including the earth matter effects \cite{Wo78,Ba80}. We place significant 
bounds in the oscillation parameter space, which  can provide useful 
guidelines for model building, for the expectations at the running 
SuperKamiokande experiment \cite{SKam}, and for the discovery potential at
future long-baseline neutrino oscillation experiments \cite{Fo96}. We also 
highlight the interplay between the results of this work and those 
obtained---within the same theoretical framework---from the analysis of
flavor oscillation searches in accelerator and reactor \cite{Sc95} or solar 
\cite{Mo96} neutrino experiments. We include a larger data set and use
a more refined statistical approach than previous three-flavor analyses 
of the atmospheric neutrino anomaly performed by other authors 
\cite{Ba88,Le88,Le93,Hk88,Mi91,Ac94,Pa94,Bi95,Go95,Na96,Ya96}
and by ourselves \cite{Sc95,Li94,Mo95}.

	The paper has the following structure. In Sec.~II we introduce 
the experimental ingredients of the analysis and describe their treatment.
In Sec.~III we briefly recall the properties of three-flavor oscillations
in the hypothesis of one dominant square mass difference. In Sec.~IV we 
perform the data analysis in the subcases of two-flavor 
$\nu_\mu\leftrightarrow\nu_\tau$ and  $\nu_\mu\leftrightarrow\nu_e$ 
oscillations. In Sec.~V we present the main results of our analysis of 
three-flavor atmospheric neutrino oscillations and discuss their relation 
with the indications coming from other (laboratory and solar) neutrino 
oscillation experiments. In Sec.~VI we summarize our work and draw our 
conclusions. In Appendices A, B, and C, we discuss respectively our 
treatment of the Kamiokande multi-GeV neutrino data, the correlation of 
the theoretical uncertainties, and the symmetry properties of
the neutrino oscillation probability.

%%%%%%%%%%%%%%%%%%%%%%%%%%%%%%%%%%%%%%%%%%%%%%%%%%%%%%%%%%%%%%%%%%%%%%%%%%%
\section{EXPERIMENTAL INGREDIENTS AND THEIR ANALYSIS}
%%%%%%%%%%%%%%%%%%%%%%%%%%%%%%%%%%%%%%%%%%%%%%%%%%%%%%%%%%%%%%%%%%%%%%%%%%%

	In this section we introduce the experimental data
and discuss briefly some technical aspects of their analysis.

%%%%%%%%%%%%%%%%%%%%%%%%%%%%%%%%%%%%%%%%%%%%%%%%%%%%%%%%%%%%%%%%%%%%%%%%%%%
\subsection{Experimental data}

	We analyze the largest set of data on the electron and muon
composition of the atmospheric neutrino flux, which can be considered 
both homogeneous and statistically consistent. In particular, we include 
the neutrino-induced $e$-like and $\mu$-like event rates measured by the 
four experiments Kamiokande \cite{Hi88,Hi92}, IMB \cite{Ca91,Be92}, 
Fr{\'e}jus \cite{Be89,Be90}, and NUSEX \cite{Ag89} at low energies 
(so-called sub-GeV data), and by the Kamiokande experiment at higher 
energies in five zenith-angle sectors \cite{Fu94} (so-called multi-GeV 
data).%
%----------------------------
\footnote{	We always include both fully and partially contained
		events in the multi-GeV Kamiokande data sample.}
%----------------------------
Concerning the multi-GeV data, we will consider either the  information 
coming from the binned angular distribution of events  (``binned 
multi-GeV''), or the reduced information coming from the angle-integrated 
number of events (``unbinned multi-GeV'').

	As it is well known, the Fr{\'e}jus and NUSEX measurements
do not confirm the Kamiokande and IMB observation of an anomalous 
muon-electron flavor composition of the atmospheric events \cite{Br92}. 
However, all the data are statistically compatible within $\sim 2\sigma$ 
or better (see Fig.~3 in \cite{Li95}) and their combination in a global 
analysis is reasonable.

	We do not include the higher-energy data pertaining only to the 
muon flavor content, the so-called upward-going muons (see, e.g., 
\cite{Fr93}). This data set has  specific experimental characteristics 
and deserves a separate theoretical analysis (in progress). However, 
we will briefly comment on upward-going muons in due course.%
%--------------
\footnote{	We note in passing that the recent reanalysis of the
 		data from the Fr{\'e}jus experiment \protect\cite{Da95}
		is relevant for the high-energy muon fluxes, but it does 
		not change significantly the previous Fr{\'e}jus results
		\protect\cite{Be89,Be90} on low-energy $e$-like and 
		$\mu$-like events.}
%--------------
We also do not include the data from the Soudan~2 experiment \cite{So95}, 
because a detailed, official analysis has not been published yet by the 
experimental collaboration.

	We emphasize that, according to the discussion in \cite{Li95}, 
we prefer to separate the electron and muon flavor information, instead
of using the popular ratio 
$R_{\mu/e}=(\mu/e)_{\rm data}/(\mu/e)_{\rm theory}$. The ratio $R_{\mu/e}$ 
allows a large cancellation of the theoretical errors \cite{Br92}, but its 
probability distribution is highly non-Gaussian.%
%--------------------
\footnote{	The ratio of two normally distributed variables obeys 
		a Cauchy distribution.}
%--------------------
We have shown \cite{Li95} that one can use only normally distributed
variables provided that, when the $\mu$ and $e$ experimental and 
theoretical rates are separated, the correlations of their uncertainties 
are included.

	In the comparison of the data with theoretical predictions, we 
use a $\chi^2$ statistic including both experimental and theoretical 
uncertainties, with  the proper error correlation matrix \cite{Li95}.

%%%%%%%%%%%%%%%%%%%%%%%%%%%%%%%%%%%%%%%%%%%%%%%%%%%%%%%%%%%%%%%%%%%%%%%%%%%
\subsection{Neutrino fluxes and interactions}

	In order to get significant constraints on the neutrino 
oscillation hypothesis, the measured numbers of $\mu$-like and $e$-like 
events produced by neutrinos and antineutrinos in each detector must be 
compared with detailed theoretical predictions.

	The theoretical calculations involve the numerical estimate of 
integrals of the kind:
%..........................................................................
\begin{equation}\label{eq:Iab}
I_{\alpha\beta}=\int\!\!d\theta \int\!\!dE_\nu\,
\frac{d^2\Phi_\alpha}{dE_\nu\,d\theta}\,
P_{\alpha\beta}
\int\!\! dE_\ell\,
\frac{d\sigma_\beta}{dE_\ell}\,
\varepsilon_{\beta}\ ,
\end{equation}
%..........................................................................
where $\theta$ is the zenith angle, $\alpha$ and $\beta$ are flavor 
indices, $E_\nu$ and $E_\ell$ are the (anti)neutrino and lepton energies,
$\Phi$ is the atmospheric (anti)neutrino flux, $P$ is the oscillation 
probability, $\sigma$ is the (anti)neutrino interaction cross-section,
and $\varepsilon$ is the lepton detection efficiency. Our estimate of 
the $I_{\alpha\beta}$'s in the multi-GeV energy range is discussed in 
Appendix A.

	Concerning the sub-GeV experiments, the ingredients $\Phi$, 
$\sigma$, and $\varepsilon$ that we use to compute  the $I_{\alpha\beta}$'s 
have been reported in our previous works \cite{Li94,Mo95}. In particular, 
we use the Bartol group calculations of $\Phi$ \cite{Ba89,Ga88} for the 
electron and muon neutrino and antineutrino fluxes at each detector 
location.

	The  Bartol fluxes  \cite{Ba89} have been used by all the four 
sub-GeV experiments in at least one simulation, and thus provide us with 
the advantages of a uniform data analysis and of a homogeneous comparison 
of our calculations with the published detector simulations.  This 
comparison has been done in  \cite{Li94,Mo95}. In particular, we refer 
the reader to Fig.~1 in \cite{Mo95}, where our absolute predictions for 
the lepton energy spectra are superposed to the corresponding published 
Monte~Carlo simulations of the Kamiokande, IMB, Fr{\'e}jus, and NUSEX 
experiments (in absence of oscillations). The agreement of our 
calculations with the published spectra is good.

	Since this work was initiated, new refined atmospheric neutrino 
flux calculations have appeared \cite{Ho95,Ag96}. These new fluxes, as far 
as we know from the published literature, have not been used yet by the 
experimental collaborations to reprocess their simulations or to reanalyze 
their data, and are not used in this work either. However, in the 
statistical analysis we conservatively associate a $\pm30\%$ uncertainty 
$(1\sigma)$ with the absolute theoretical neutrino fluxes. This error 
accounts conservatively for the spread in the atmospheric neutrino 
calculations published so far \cite{Li95}. The uncertainties of the $\mu$ 
and $e$ neutrino fluxes  are highly correlated  (see \cite{Li95} and 
Appendix~B for more details). Such correlation  is even more important 
than the absolute magnitude of the errors themselves in driving the fits
to atmospheric $\nu$ data.  In fact, it will be shown that a reduction of 
the flux error from our default value, $\sigma_{\rm flux}=30\%$, to 
$\sigma_{\rm flux}=20\%$ or even $15\%$ does not change significantly the 
results of the oscillation fits. In the absence of oscillations, it has 
already been shown in Ref.~\cite{Li95} (Table~III) that the statistical 
significance of the anomaly does not change much by reducing the flux 
error from $30\%$ to $20\%$.

	The last ingredient of Eq.~(\ref{eq:Iab}) to be discussed is the 
oscillation probability $P_{\alpha\beta}$. This is the subject of the next 
Section.

%%%%%%%%%%%%%%%%%%%%%%%%%%%%%%%%%%%%%%%%%%%%%%%%%%%%%%%%%%%%%%%%%%%%%%%%%%%
\section{THEORETICAL FRAMEWORK}
%%%%%%%%%%%%%%%%%%%%%%%%%%%%%%%%%%%%%%%%%%%%%%%%%%%%%%%%%%%%%%%%%%%%%%%%%%%

	The calculation of the atmospheric neutrino oscillation
probabilities requires a well-defined theoretical framework. The 
three-flavor framework used in this paper is completely specified 
by the neutrino spectrum shown in Fig.~1.

	In Fig.~1 we show the adopted spectrum of neutrino mass 
eigenstates $(\nu_1,\,\nu_2,\,\nu_3)$. Two states $(\nu_1,\,\nu_2)$
are assumed to be almost degenerate in mass. The third state, $\nu_3$,
is largely separated in mass from the almost degenerate doublet, with
$|m^2_3-m^2_{1,2}|\simeq m^2$ being the dominant square mass difference 
driving the atmospheric neutrino oscillations. This situation can be 
realized either in scenario (a) or in scenario (b) of Fig.~1, that is, 
with $\nu_3$ either lighter or heavier than $\nu_1$ and $\nu_2$.

	In both scenarios of Fig.~1, the subdominant square mass difference 
between $\nu_2$ and $\nu_1$, $\delta m^2=m^2_2-m^2_1$, is assumed to be 
too small to produce detectable effects in the energy range explored by 
current atmospheric neutrino experiments. This assumption holds, for 
instance, if the parameter $\delta m^2$ is used to fit solar neutrino 
data \cite{Mo96}  ($\delta m^2 \simeq 5\times 10^{-6}$ eV$^2$ at the 
best-fit point for matter-enhanced oscillations). At the scale of the
atmospheric neutrino experiments, we simply set $\delta m^2=0$. The limits 
of this approximations are commented in Sec.~V~C. The same approximation 
has been used in \cite{Sc95} where we studied accelerator and reactor 
neutrino oscillations, and to which we refer the reader for further 
details and references. We collectively consider the accelerator, reactor, 
and atmospheric neutrino experiments as ``terrestrial'' oscillation
experiments.

	At zeroth order in $\delta m^2/m^2$ the two scenarios of Fig.~1 
are physically different for atmospheric neutrinos propagating in the 
earth matter (see Appendix C), while they are not distinguishable either by 
means of accelerator and reactor neutrino oscillation searches (in vacuum) 
\cite{Sc95} or by solar neutrino oscillations (in vacuum or in matter) 
\cite{Mo96}. In this sense, atmospheric neutrinos potentially provide a 
unique information on the neutrino spectrum.

	Notice that the spectra (a) and (b) in Fig.~1 are simply related by: 
%.........................................................................
\begin{equation}
{\rm (a)}\to{\rm (b)}\Longleftrightarrow +m^2\to-m^2\ .
\end{equation}
%.........................................................................
In the following, we will explicitly distinguish cases (a) and (b) 
whenever necessary.

	In the framework characterized by the spectrum of Fig.~1 one has
two important simplifications with respect to the most general
three-flavor oscillation scenario (see  \cite{Sc95} and Appendix C): 
(1) 	neutrino mixing in terrestrial oscillation experiments  can be 
	described by just two mixing angles, $\psi$ and $\phi$; and
(2) 	effects related to the a possible CP violating phase are 
	unobservable.
It follows that the parameter space for terrestrial (including atmospheric)
three-flavor neutrino oscillations can be described in terms of three 
parameters: $(m^2,\,\psi,\,\phi)$.

	The angles $\psi$ and $\phi$ are defined as in \cite{Ku87}. In the
standard parametrization of the mixing matrix \cite{Mo94} the following 
identifications hold:
%..........................................................................
\begin{equation}
		\psi=\theta_{23}, \  \phi=\theta_{13}\ .
\end{equation}
%..........................................................................

	The three-flavor framework with one dominant square mass difference 
is the simplest  extension of the two-generation formalism in which all the 
two-flavor oscillation channels are open.  The subcases of pure two-flavor 
oscillations $\nu_e\leftrightarrow\nu_\tau$, $\nu_\mu\leftrightarrow\nu_e$,
and $\nu_\mu\leftrightarrow\nu_\tau$, are recovered in the limits $\psi=0$, 
$\psi=\pi/2$, and $\phi=0$ respectively.  A clear graphical representation 
of the general three-flavor case and of its two-flavor limits is provided
by Fig.~2 in \cite{Sc95}.  Here we do not pay attention to the 
$\nu_e\leftrightarrow\nu_\tau$ subcase, since it does not solve---but 
rather aggravates---the atmospheric neutrino anomaly. We will briefly 
comment on a theoretically interesting, genuine three-flavor subcase of 
this framework, the so-called threefold maximal mixing scenario 
\cite{Ki95,Ha95}, in Sec.~V. In our notation, threefold maximal mixing 
corresponds to $(\tan^2\psi,\,\tan^2\phi)=(1,1/2)$ at any $m^2$.

	With a neutrino mass  spectrum as in  Fig.~1, the calculation of 
the vacuum oscillation probability for neutrinos arriving from above the 
horizon is straightforward (see Appendix~C). However, the vacuum 
approximation is not adequate for neutrinos which travel in the earth 
matter for a large fraction of their path length. For these neutrinos one 
expects substantial deviations from the vacuum oscillation probability  
\cite{Da84,Ay84,Ca86,Au87,Mi90,Pa94} when  
$m^2\sim 2\sqrt{2} G_F {\overline N}_e E_\nu$, where ${\overline N}_e$ is 
the typical electron density along the  neutrino trajectory. This 
implies significant matter effects in the range 
$m^2\sim 10^{-4}$--$10^{-3}$ eV$^2$ for sub-GeV observables
and in the range
$m^2\sim 10^{-3}$--$10^{-2}$ eV$^2$ for multi-GeV observables.%
%----------------------------------
\footnote{	It should be noted, however, that matter effects do not 
		totally disappear even in the limit $m^2\to\infty$, see 
		Appendix C.}
%-----------------------------------
Moreover, a genuine three-flavor effect takes place \cite{Pa94}: the 
effective mass eigenstates $\nu'_{1,2}$ in matter are not exactly
degenerate, and the phase variation associated to their splitting
is relevant for path lengths comparable to the earth radius. This 
additional phase disappears in the two-flavor subcases, since one of the 
``degenerate'' neutrinos decouples.

	We take into account the earth matter effects by solving 
numerically the neutrino propagation equations with an assigned earth 
density profile \cite{An89}. In order to save computer time, the density 
profile is modeled as a 5-step function \cite{Mo95} with steps 
corresponding to the five relevant radial shells. This approximation is 
sufficiently accurate for our purposes.

	We have a  final remark on the graphical representations of the 
results. In \cite{Sc95} we have shown that the three-flavor parameter
space for terrestrial neutrino oscillations can be usefully charted using 
the (logarithmic) coordinates $(m^2,\,\tan^2\psi,\,\tan^2\phi)$.
In the following, the results of our analysis will be displayed in several 
plane sections of the space $(m^2,\,\tan^2\psi,\,\tan^2\phi)$ at fixed 
values of either $m^2$, $\phi$, or $\psi$.

%%%%%%%%%%%%%%%%%%%%%%%%%%%%%%%%%%%%%%%%%%%%%%%%%%%%%%%%%%%%%%%%%%%%%%%%%%%
\section{TWO-FLAVOR ANALYSIS}
%%%%%%%%%%%%%%%%%%%%%%%%%%%%%%%%%%%%%%%%%%%%%%%%%%%%%%%%%%%%%%%%%%%%%%%%%%%

	In this section we show the results of our analysis in the subcases 
of pure $\nu_\mu\leftrightarrow\nu_\tau$ and  $\nu_\mu\leftrightarrow\nu_e$ 
two-flavor oscillations. The discussion of these cases is interesting in 
itself, and helps in understanding the more complicated situation of 
three-flavor oscillations.

%%%%%%%%%%%%%%%%%%%%%%%%%%%%%%%%%%%%%%%%%%%%%%%%%%%%%%%%%%%%%%%%%%%%%%%%%%%
\subsection{Pure $\nu_\mu\leftrightarrow\nu_\tau$ oscillations}

	In our framework we recall that the subcase of pure
$\nu_\mu\leftrightarrow\nu_\tau$ oscillations is reached in the limit 
$\phi\to 0$, which leaves  $(m^2,\,\psi)$ as relevant variables. 
This subcase is particularly simple from a theoretical viewpoint, 
since  $\nu_e$ is decoupled and thus the earth matter does not affect the
oscillations. As a consequence (see Appendix~C), the scenarios (a) and (b)
in Fig.~1 become physically equivalent, and the formalism is invariant 
under the substitution $\psi\to\frac{\pi}{2}-\psi$. One usually takes 
advantage of this symmetry to restrict the range of the mixing angle 
$\psi$ to $[0,\,\pi/4]$. Here we show the full range $\psi\in[0,\,\pi/2]$ 
in order to mark the difference with the  $\nu_\mu\leftrightarrow\nu_e$ 
case (Sec.~IV~B), where the symmetry $\psi\to\frac{\pi}{2}-\psi$ is broken
by matter effects.

	In Fig.~2 we show the results of our analysis as bounds in the 
$(\tan^2\psi,\,m^2)$ plane at $90\%$ C.L.\ (solid lines) and $99\%$ C.L.\ 
(dotted lines), which corresponds to $\Delta\chi^2=4.61$ and 9.21 
respectively ($N_{\rm DF}=2$). Notice the mirror symmetry of all the 
curves  with respect to the axis $\psi=\pi/4$.

	To avoid ambiguities, the allowed regions are marked by  stars.
We refrain from showing the best-fit points in each panel of Fig.~2, since 
the $\chi^2$ is often relatively flat around the minimum (except for the 
combination of several data), and the best-fit point is not particularly 
informative. We give the best-fit coordinates only for the combination of 
all data.

	In the first four panels of Fig.~2 (left to right, top to bottom)
we display  the parameter regions individually allowed by the four sub-GeV 
experiments, Fr{\'e}jus, NUSEX, IMB, and Kamiokande. The Fr{\'e}jus and 
NUSEX experiments are consistent with no oscillations, and thus strongly
disfavor the situation of maximal mixing $(\psi\simeq\pi/4)$, at least 
for not too small $m^2$ where they are not sensitive to oscillations.
The IMB and Kamiokande experiments are instead compatible with large 
mixing in a wide range of $m^2$ ($m^2\gtrsim 10^{-4}$ eV$^2$), although 
the best fit is not reached exactly for maximal mixing.

	It should be noted that the sub-GeV experiments have still some 
sensitivity to values of $m^2$ as small as $10^{-4}$ eV$^2$ 
(see, e.g., \cite{Ba88,Go95,Mo95}). For instance, the neutrino phase 
variation due to oscillations can be of ${\cal O}(1)$ for $E_\nu=1$ GeV,  
$m^2=10^{-4}$ eV$^2$, and a path length equal to the earth diameter. 
This fact, combined with our conservative error estimates, explains the 
extension of the IMB and  Kamiokande allowed regions down to values as 
small as $m^2\simeq 10^{-4}$ eV$^2$ in Fig.~2. Our Kamiokande allowed 
regions are larger than those derived by the Kamiokande collaboration  
analysis \cite{Fu94}. However, one has to consider that we use
only the published experimental information, while the Kamiokande 
collaboration uses energy-angle lepton distributions that are not 
published, and also adopts a  rather different approach in the statistical 
analysis of the data \cite{Fu94}.

	The 5th and 6th panels in Fig.~2 show the constraints from the
multi-GeV Kamiokande data, taken both unbinned and binned respectively. 
The sensitivity to small $m^2$ is lower than in the sub-GeV cases, since 
the average  neutrino energy is higher.  In the binned case (full 
information from the angular distribution), $m^2$ is bounded from above 
at $90\%$ C.L. In fact, the measured angular distribution is inconsistent 
with the flat oscillation probability corresponding to averaged fast 
oscillations ($m^2\to\infty$) \cite{Fu94}.

	In the 7th and 8th panels of Fig.~2, we combine the sub-GeV and 
multi-GeV (unbinned and binned) Kamiokande data.  In the 9th panel all 
sub-GeV data are combined together. Notice that maximal mixing 
$(\psi\simeq\pi/4)$ is strongly disfavored for $m^2\gtrsim 10^{-2}$ eV$^2$, 
due to the inclusion of the Fr{\'e}jus and NUSEX data in the fit. In the 
10th panel we combine the sub-GeV data with the multi-GeV unbinned data. 
The multi-GeV unbinned data disfavor the lowest values of $m^2$ that would 
be allowed by the sub-GeV data alone.

	In the 11th panel we combine  sub-GeV and binned multi-GeV data. 
Therefore, this panel contains the maximum information from the atmospheric 
neutrino experiments (8 sub-GeV + 10 multi-GeV observables). The best-fit 
is reached at  $(m^2,\,\tan^2\psi)\simeq(5\times10^{-3}{\rm\ eV}^2,\,0.63)$ 
and at the symmetric point
$(m^2,\,\tan^2\psi)\simeq(5\times10^{-3}{\rm\ eV}^2,\,1/0.63)$.  The 
best-fit value of $m^2$ is somewhat lower than the popular value
$10^{-2}$ eV$^2$, since we include the Fr{\'e}jus and NUSEX data. These 
data are compatible with $m^2=0$ (no oscillation) and thus tend to drag
the $m^2$-fit to lower values than those favored by IMB and Kamiokande 
alone. The value of the absolute $\chi^2$ at the minimum is 19.7, which 
represents a good fit to the 18 atmospheric observables (with the freedom 
to vary only $(m^2,\,\psi)$).  One should compare this value with the 
corresponding (worse) fit in the no-oscillation hypothesis,
$\chi^2_{\rm no\ osc}=44.6$.

	The last panel in Fig.~2 shows the region allowed by the 
combination of all the established accelerator oscillation searches. The 
region allowed at $90\%$ by all atmospheric neutrino data is also allowed 
by present accelerator data (reactor data place no bounds in the
$\nu_\mu\leftrightarrow\nu_e$ oscillation limit). The atmospheric  bounds 
are in conflict with the accelerator bounds  only at  $\sim99\%$ C.L.\ 
and high $m^2$.

	In Fig.~2, the magnitude of the theoretical neutrino flux error,
$\sigma_{\rm flux}$, has been taken equal to $30\%$ (default value). 
This choice is not decisive in the fit, as shown in Fig.~3.

	In Fig.~3 we compare two representative fits  (all sub-GeV data, 
and  all sub-GeV~+~multi-GeV data)  using both $\sigma_{\rm flux}=30\%$ 
(solid contours) and  $\sigma_{\rm flux}=15\%$ (dotted contours). The 
differences between the two cases are  very small. In fact, since we 
include the proper correlations between the neutrino fluxes, the $\nu_e$ 
and $\nu_\mu$ flux errors nearly ``cancel'' in the analysis, with a 
residual $\pm 5\%$ difference allowed in the relative $\nu_\mu/\nu_e$  
flux normalization. Of course, a similar cancellation is reached by using 
the ratio  $R_{\mu/e}=(\mu/e)_{\rm data}/(\mu/e)_{\rm theory}$, but with 
the  serious disadvantage of obtaining a highly non-gaussian distribution
for $R_{\mu/e}$ \cite{Li95}.  Previous analyses of the atmospheric neutrino 
anomaly that applied gaussian statistics to  $R_{\mu/e}$, including our 
works \cite{Li94} and \cite{Mo95}, may thus have overestimated the 
statistical significance of the anomaly, and underestimated the 
mass-mixing regions allowed by the oscillation hypothesis.

	In this work we have not analyzed the so-called upward-going muon 
data. These data are essentially consistent with no oscillations 
\cite{Fr93}, although with large, non-cancelable flux errors---there are 
no ``upward-going electrons'' to be used for comparison. Thus we expect 
that, in a combined analysis: 
(1) 	the upward-going muon data should have a smaller statistical 
	weight than the atmospheric data considered here, which should 
	drive the fit; and 
(2) 	the inclusion of upward muon data should anyway disfavor a too 
	strong suppression of the muon rates. 
This implies that the fit  discussed in this section  should be generally 
worsened for nearly maximal mixing ($\psi\sim \pi/4$). The three-flavor
analysis of upward-going muon data is in progress and will be presented 
in a separate paper.

%%%%%%%%%%%%%%%%%%%%%%%%%%%%%%%%%%%%%%%%%%%%%%%%%%%%%%%%%%%%%%%%%%%%%%%%%%%
\subsection{Pure $\nu_\mu\leftrightarrow\nu_e$ oscillations}

	We recall that, in our framework, the subcase of pure
$\nu_\mu\leftrightarrow\nu_e$ oscillations is reached in the limit 
$\psi\to \pi/2$, which leaves $(m^2,\,\phi)$ as relevant variables.

	This limit is  more complicated than the 
$\nu_\mu\leftrightarrow\nu_\tau$ limit, since matter effects are not 
decoupled. It  follows that the intervals $\phi\in[0,\,\pi/4]$ and 
$\phi\in[\pi/4,\,\pi/2]$ are not physically equivalent, and  the 
scenarios (a) and (b) shown in Fig.~1 are also not equivalent.  However, 
in the $\nu_\mu\leftrightarrow\nu_e$ oscillation case there is an 
interesting symmetry:  the physics in (b) is equivalent to the physics 
in (a), provided that $\phi$ is replaced by its complementary angle 
($\phi\to\frac{\pi}{2}-\phi$). This symmetry (discussed in Appendix~C) 
allows us to consider only one scenario, (a) for definiteness.

	In Fig.~4, we show the results of our $\nu_\mu\leftrightarrow\nu_e$
oscillation analysis in scenario (a). The  corresponding results in 
scenario (b) are obtained by looking at the same figure in a mirror.
Since the panels in Fig.~4  ($\nu_\mu\leftrightarrow\nu_e$ case) are 
analogous to the panels in Fig.~2 ($\nu_\mu\leftrightarrow\nu_\tau$ case)
we just highlight the differences between the results in Figs.~4 and 2.

	At large $m^2$, the maximal mixing value $\phi\simeq\pi/4$ is in 
general  disfavored, since ``too many'' $\nu_\mu$'s oscillate into 
$\nu_e$'s, the muon (electron) rate become too suppressed (enhanced), and 
thus the flavor anomaly is overbalanced by the oscillations. In Fig.~2 
this situation was also disfavored, but not as strongly, since in the 
$\nu_\mu\leftrightarrow\nu_\tau$  case the electron rates are unaffected 
by oscillations.

	At small $m^2$, the contours are affected by the earth matter 
effects that are also responsible for the asymmetry with respect to the 
axis $\phi=\pi/4$. Notice that, in general,  values of $m^2$ close to
$10^{-4}$ eV$^2$ are more disfavored than in Fig.~2.

	The combination of all data is shown in the 11th panel of Fig.~4
(all sub-GeV and multi-GeV binned). The allowed region should be compared 
with the constraints coming from accelerator and reactor searches. These 
constraints essentially exclude the upper part of the region preferred by 
atmospheric $\nu$ data, but are compatible with the (larger) lower part.

	The best-fit for the combination of all data is obtained for 
$(m^2,\,\tan^2\phi)=(6.6\times10^{-3}{\rm\ eV}^2,\,0.36)$. For scenario (b),
the best-fit would be obtained at the symmetric point 
$(m^2,\,\tan^2\phi)=(6.6\times10^{-3}{\rm\ eV}^2,\,1/0.36)$. The value
of $\chi^2$ at the minimum is  $\chi^2_{\rm min}=15.6$, 
about 4 units lower than in the case of $\nu_\mu\leftrightarrow\nu_\tau$ 
oscillations $(\chi^2_{\rm min}=19.7)$. Therefore 
$\nu_\mu\leftrightarrow\nu_e$ oscillations 
appear to be  preferred to $\nu_\mu\leftrightarrow\nu_\tau$ oscillations 
in our global analysis.

	The preference for the $\nu_\mu\leftrightarrow\nu_e$ case is 
essentially driven by the multi-GeV data. In Fig.~5, we show the measured
and expected $\mu$-like and $e$-like event rates  in the five zenith-angle 
bins used by the Kamiokande collaboration for the multi-GeV data. The 
$\mu$ and $e$ rates have been conventionally  divided by their theoretical 
(central) value in absence of oscillations, $\mu_0$ and $e_0$ (see also 
Ref.~\cite{Li95} where we introduced this graph). The ellipses represent
$1\sigma$ contours ($\Delta\chi^2=1$).

	The first of the horizontal  panels in Fig.~5 refers to the 
no-oscillation case, which is clearly a bad fit to the data. The second 
panel refers to the best-fit to multi-GeV data with pure 
$\nu_\mu\leftrightarrow\nu_\tau$ oscillations. In this case, only the 
theoretical muon rates can vary, and the fit is not particularly
good, although certainly better than in the no-oscillation case. The 
third panel refers to pure $\nu_\mu\leftrightarrow\nu_e$ oscillations.
In this case, both $\mu$ and $e$ rates vary with oscillations, and one 
can get higher theoretical electron rates that match better the 
experimental data (notice in particular the good fit in the first bin).
The fourth panel refers to the general case of three-flavor oscillations,
with $(m^2,\,\psi,\,\phi)$  unconstrained. The $3\nu$ best-fit is only 
slightly better than in the $\nu_\mu\leftrightarrow\nu_e$ case. The 
differences between the overall fits in the third and  fourth panels
are only appreciable numerically and not by eye. In conclusion,
$\nu_\mu\leftrightarrow\nu_e$ oscillations seem to provide a 
close-to-optimal fit to multi-GeV data.

	In Fig.~6 we illustrate the importance of including the earth 
electron density in  $\nu_\mu\leftrightarrow\nu_e$ oscillations. 
Figure~6 is analogous to  Fig.~4 but {\em without\/} matter effects. The 
allowed regions  in Fig.~6 and Fig.~4 differ considerably for low $m^2$;  
when matter effects are  included (Fig.~4) the lowest values of $m^2$ do 
not provide a good fit \cite{Ak93}. The reason is that for not too small 
$\nu_\mu\leftrightarrow\nu_e$ mixing (i.e., for $\tan^2\phi\sim0.1$--10), 
the mixing angle in matter is rapidly  suppressed for $m^2\to 0$ 
[see Eq.~(C4)], and so are the oscillations that should solve the 
flavor anomaly.

%%%%%%%%%%%%%%%%%%%%%%%%%%%%%%%%%%%%%%%%%%%%%%%%%%%%%%%%%%%%%%%%%%%%%%%%%%%
\section{THREE-FLAVOR ANALYSIS in the 
$({\lowercase{m}}^2,\,\tan^2\psi,\,\tan^2\phi)$ PARAMETER SPACE}
%%%%%%%%%%%%%%%%%%%%%%%%%%%%%%%%%%%%%%%%%%%%%%%%%%%%%%%%%%%%%%%%%%%%%%%%%%%

	In this section we show the results of our analysis of atmospheric
neutrino data within the three-flavor framework discussed in  Sec.~III. 
The free parameters of the fit are  $(m^2,\,\psi,\,\phi)$. The analysis
includes all the sub-GeV data and the binned multi-GeV data. We represent 
the results in the mixing-mixing plane $(\tan^2\psi,\,\tan^2\phi)$ at 
representative values of $m^2$, in both scenarios (a) and (b) of Fig.~1.

%%%%%%%%%%%%%%%%%%%%%%%%%%%%%%%%%%%%%%%%%%%%%%%%%%%%%%%%%%%%%%%%%%%%%%%%%%%
\subsection{Scenario (a)}

	In Fig.~7 we show the results of the fit to all the atmospheric 
data in scenario (a). The solid (dotted) lines represent sections, at 
given values of $m^2$,  of the three-dimensional 
$(m^2,\,\tan^2\psi,\,\tan^2\phi)$ manifold allowed by the fit at 
$90\%$ C.L.\ ($99\%$ C.L.)  for $N_{\rm DF}=3$ ($\Delta\chi^2=6.25$ and 
11.34 respectively).

	The representative values of $m^2$ range from $0.18$ eV$^2$ down 
to $3.2\times 10^{-4}$ eV$^2$. For $m^2\gtrsim 10^{-2}$ eV$^2$ (first six 
panels),  reactor and accelerator neutrino oscillation experiments also 
place bounds on the mixing angles. These bounds are discussed separately 
in Sec.~V~C.

	We recall that in each panel of Fig.~7 the right-hand side 
corresponds (asimptotically) to the limit of pure 
$\nu_\mu\leftrightarrow\nu_e$ oscillations, and the lower side to pure 
$\nu_\mu\leftrightarrow\nu_\tau$  oscillations. The left-hand side, 
corresponding to pure  $\nu_e\leftrightarrow\nu_\tau$ oscillations, 
never represents an acceptable fit to the data. The asymptotic regime is 
already reached at the ends of the $\tan^2\psi$ and   $\tan^2\phi$ ranges 
adopted in Fig.~7.

	The best three-flavor fit is reached at 
$(m^2,\,\tan^2\psi,\,\tan^2\phi)=
(4.6\times10^{-3}{\rm\ eV}^2,\,7.07,\,0.28)$.
The corresponding value of the $\chi^2$ is $\chi^2_{\rm min}=14.8$, which
represents a good fit to the 18 atmospheric observables, given the freedom 
of varying the three parameters  $(m^2,\,\psi,\,\phi)$.

	At $90\%$ C.L.\ there are both an upper and a lower bound on
$m^2$: $0.6\times 10^{-3}\lesssim m^2\lesssim 1.5\times 10^{-1}$ eV$^2$.
The upper bound,  provided by the inclusion of multi-GeV data, however, 
disappears at $\sim 95\%$ C.L.\ (see below). The inclusion of laboratory 
oscillation data would make the upper bound tighter  
($m^2\lesssim 6\times 10^{-2}$ eV$^2$ at $90\%$ C.L., see Sec.~V~C).

	For relatively large $m^2$ ($m^2\gtrsim 2\times 10^{-2}$ eV$^2$)
the situations of maximal $\nu_\mu\leftrightarrow\nu_e$ mixing
	$[(\tan^2\psi,\,\tan^2\phi)=(\infty,\,1)]$, 
of maximal $\nu_\mu\leftrightarrow\nu_\tau$ mixing
	$[(\tan^2\psi,\,\tan^2\phi)=(1,\,0)]$,
and of threefold maximal mixing 
	$[(\tan^2\psi,\,\tan^2\phi)=(1,\,1/2)]$,
are not allowed. All these twofold and threefold maximal mixing situations 
are allowed, however, in the range  
$1.5\times10^{-3}\lesssim m^2\lesssim 7\times 10^{-3}$ eV$^2$ (at least).
In this range, the vacuum oscillation probabilities for threefold
maximal mixing \cite{Ki95,Ha95} get significant corrections when
matter effects are included (see Appendix~C).

	In many panels of Fig.~7, the allowed region interpolates smoothly
between the two-flavor oscillation limits $\nu_\mu\leftrightarrow\nu_\tau$ 
and $\nu_\mu\leftrightarrow\nu_e$ \cite{Hk88,Mi91,Ac94,Pa94,Sc95}. 
However, for  $m^2\gtrsim 2\times 10^{-2}$ eV$^2$, pure 
$\nu_\mu\leftrightarrow\nu_\tau$ oscillations are disfavored since the 
global fit improves towards the $\nu_\mu\leftrightarrow\nu_e$ oscillation 
limit.

	The limits on $m^2$ for unconstrained $\phi$ and $\psi$ are
particularly interesting as guidelines for future long-baseline neutrino 
oscillation searches. In Fig.~8 we thus show the value of $\Delta\chi^2$ as 
a function of $m^2$ only ($\psi$ and $\phi$ are projected away). The solid 
line refers to the default flux error $(\sigma_{\rm flux}=30\%)$. The 
dashed line, which refers to  $\sigma_{\rm flux}=20\%$, is not 
significantly different from the solid line.

	From Fig.~8 one can trace the upper and lower bounds on $m^2$ 
placed by all the atmospheric neutrino data at any given C.L. However, 
for $m^2\to\infty$  the $\Delta\chi^2$ tends to the asymptotic limit 
$\sim7$ (not shown). It follows that, in the adopted  three-flavor 
framework, atmospheric neutrinos place no upper bound on $m^2$ at 
$95\%$ C.L. $(N_{\rm DF}=3)$.  Atmospheric neutrino data would also 
not place any upper bound on $m^2$ if the zenith-angle dependence of the 
multi-GeV data were discarded, i.e.\ if {\em unbinned\/} multi-GeV data
were used in the fit.

%%%%%%%%%%%%%%%%%%%%%%%%%%%%%%%%%%%%%%%%%%%%%%%%%%%%%%%%%%%%%%%%%%%%%%%%%%%
\subsection{Scenario (b)}

	In Fig.~9 we show the results of the fit to all the atmospheric 
data in scenario (b) of Fig.~1.  Figure~9 is analogous to Fig.~7, but all 
the calculations have been  done with $-m^2$ instead of $+m^2$.
The solid (dotted) lines represent sections, at given values of $-m^2$,
of the three-dimensional $(m^2,\,\tan^2\psi,\,\tan^2\phi)$ manifold 
allowed by the data at $90\%$ C.L.\ ($99\%$ C.L.)  for $N_{\rm DF}=3$.

	In scenario (b), the best-fit point is  
$(m^2,\,\tan^2\psi,\,\tan^2\phi)=
(6.8\times10^{-3}{\rm\ eV}^2,\,11.2,\,2.82)$. The corresponding  value of 
$\chi^2_{\rm min}$ is 15.1, which is almost as good as in scenario (a).

 	As expected from symmetry arguments (see Appendix C),  Fig.~9 and 
Fig.~7 coincide in the limit of pure $\nu_\mu\leftrightarrow\nu_\tau$ 
oscillations (lower side of each panel).  In the limit of pure
$\nu_\mu\leftrightarrow\nu_e$ oscillations (right side of each panel)
these figures coincide%
%--------------------------------
\footnote{	The coincidence of the C.L.\  contours in these two-flavor 
		limits is not perfectly realized because the best-fit point
		and the value of the $\chi^2$ at the minimum are not 
		exactly equal (and are not expected to be equal)
		in scenarios (a) and (b).}
%--------------------------------
modulo the replacement  $\phi\to\frac{\pi}{2}-\phi$. In the intermediate, 
genuine three-flavor mixing cases, the bounds shown in Figs.~9 and 7 are 
slightly different at any $m^2$.

	The differences between the three-flavor fits in Figs.~7 and 9 are 
not unexpected, since they correspond to two physically different 
scenarios. Unfortunately, the differences are quite small, implying that 
the available information on atmospheric neutrinos is not
 sufficiently accurate to discriminate the two cases (a) and (b). 
A significant discrimination would have important implications.
For instance a hypothetical, pronounced preference of atmospheric data
for scenario (a) would support the theoretical prejudice that the spectrum
of neutrino masses is similar to the spectrum of charged fermions 
(two light states and a third, much heavier state). It will be interesting 
to see if the  atmospheric neutrino data that are being collected
with high statistics by the running SuperKamiokande experiment will show 
a preference for one of the two scenarios (if they confirm the flavor 
anomaly). We recall that this  information cannot be provided either by
accelerator,%
%--------------------------------------
\footnote{	However, futuristic accelerator oscillation searches with 
		extremely long baselines (greater than $10^3$ km) could 
		in principle probe the difference between scenarios (a)
		and (b) through earth matter effects.}
%--------------------------------------
reactor, or solar neutrino oscillation searches, i.e., these experiments 
{\em a priori\/} do not distinguish  the scenarios (a) and (b) at
zeroth order in $\delta m^2/m^2$.

	Finally,  we complete our survey of the fit in scenario (b) by 
showing in Fig.~10 the dependence of $\Delta\chi^2$ on $-m^2$. Fig.~10 
is the analogous to Fig.~8 in scenario (b).

%%%%%%%%%%%%%%%%%%%%%%%%%%%%%%%%%%%%%%%%%%%%%%%%%%%%%%%%%%%%%%%%%%%%%%%%%%%
\subsection{Comparison with other oscillation searches}

	The three-flavor bounds  shown in Figs.~7 and 9 and discussed in 
Sections V~A and V~B were obtained by fitting only the atmospheric
neutrino data. In this section, we show their interplay with the 
independent constraints obtained  by  accelerator and reactor neutrino 
oscillation searches \cite{Sc95} and by solar neutrino experiments 
\cite{Mo96}.

	The analyses \cite{Sc95} and \cite{Mo96} were performed under the 
same assumption on the neutrino spectrum shown in Fig.~1, namely that the
two independent neutrino mass squared differences, 
$\delta m^2=|m^2_2-m^2_1|$ and $m^2=|m^2_3-m^2_2|$, are largely
separated: $\delta m^2\ll m^2$.  Accelerator and reactor neutrinos were 
assumed to probe, as the atmospheric neutrinos, the dominant square mass
difference $m^2$, as the slow oscillations driven by  $\delta m^2$
were effectively frozen. Conversely, solar neutrinos were assumed to probe 
the subdominant square mass difference $\delta m^2$, as the fast 
oscillations driven by $m^2$ were effectively averaged out. The parameter 
spaces probed by terrestrial (accelerator, reactor, and atmospheric) 
neutrinos and by solar neutrinos have been discussed thoroughly 
in \cite{Sc95} and \cite{Mo96}.

	In Fig.~11, we show the bounds coming from the established
accelerator and reactor oscillation searches \cite{Sc95} (the recent data 
from the Liquid Scintillator Neutrino Detector (LSND) experiment 
\cite{LNS2} are not included). These negative searches  exclude horizontal 
bands in the $(\tan^2\psi,\,\tan^2\phi)$ plane for $m^2\gtrsim 10^{-2}$ 
eV$^2$ ($90\%$ C.L.\ limits are shown). Superposed are the $90\%$ C.L.\
regions from the atmospheric neutrino analysis (only in scenario (a) for 
definiteness). The $\nu_\mu\leftrightarrow\nu_e$ oscillation limit (right 
side of the panels) is generally disfavored in the range probed by 
laboratory oscillation experiments,  mainly because it is not consistent 
with the unsuccessful $\nu_e$ disappearance searches at reactors. For 
$m^2\gtrsim 6\times 10^{-2}$ eV$^2$ the atmospheric data are not 
compatible with existing laboratory limits at any $\psi$ or $\phi$. For 
$m^2\lesssim 10^{-2}$ eV$^2$, however, there are no laboratory constraints 
on neutrino mixing, and the fits to atmospheric data (six lowest panels of 
Figs.~7 and 9) are unaffected.

	In conclusion, the limits placed by the combination of 
accelerator, reactor, and atmospheric neutrino data on $m^2$ in our 
three-flavor framework are 
$6\times 10^{-4}{\rm\ eV}^2 \lesssim m^2 \lesssim 6\times 10^{-2}$ eV$^2$
($90\%$ C.L., $N_{\rm DF}=3$). Future long-baseline experiments  will be 
able to explore a large fraction of this $m^2$ range.

	Concerning solar neutrinos, we have emphasized in \cite{Mo96}
that they probe the same mixing angle $\phi$ probed by terrestrial
(atmospheric, accelerator and reactor) neutrino experiments. If one 
accepts the explanation of solar neutrino deficit provided by
matter-enhanced oscillations, then the data constrain $\phi$ in the 
range $\tan^2\phi\lesssim1.4$ at $90\%$ C.L. $(N_{\rm DF}=3)$, with a 
preference for the value $\tan^2\phi=0$ \cite{Mo96}.  These bounds 
coming from solar $\nu$ data exclude  significant parts of the  
large-$\phi$ regions allowed by atmospheric neutrino data in Figs.~4, 
7, 9, and 11. This  should be taken into account when building models 
of neutrino masses and mixings which try to accommodate both the solar 
neutrino deficit and the atmospheric neutrino anomaly.

	In our framework, once $\delta m^2$ is fixed by  solar neutrinos, 
it is not possible to explain both the atmospheric neutrino anomaly and 
the recent LSND evidence \cite{LNS2} for oscillations with the remaining 
mass parameter $m^2$.  We have shown previously that atmospheric neutrino 
data alone place an upper limit to $m^2$ of about $1.5\times 10^{-1}$ 
eV$^2$, which is strengthened  to $m^2\lesssim 6\times 10^{-2}$ eV$^2$ 
when accelerator and reactor data are included. The range 
$m^2\lesssim 6\times 10^{-2}$ eV$^2$ is too low for significant neutrino 
oscillation effects at LSND. It follows that, in our framework, one can 
either fit the atmospheric neutrino anomaly or the LSND event excess, but 
not both. We intend to examine the second option (fit to LSND data) in a
separate publication; some interesting results were  already obtained 
with older LSND data in \cite{Sc95} (see there Figs.~11 and 12 and the 
related discussion). However, it should be noted that the atmospheric 
data fit at large $m^2$ is essentially driven by the angular distribution 
of multi-GeV events observed in Kamiokande.  If this information were 
discarded, i.e., if one used only unbinned multi-GeV data, then atmospheric
neutrino data alone would not place an upper bound on $m^2$, and one could
find \cite{Ca96} a very small region of the parameter space which is 
marginally allowed by both LSND and sub-GeV atmospheric data, as well by
present accelerator and reactor constraints. This solution is admittedly
fragile \cite{Ca96}.

	A final remark is in order. The basic assumption  underlying this 
work and Refs.~\cite{Sc95,Mo96} is that $\delta m^2\ll m^2$ (Fig.~1). 
Then all the calculations are done at zeroth order in $\delta m^2/ m^2$, 
i.e.,  one takes $\delta m^2=0$ and $m^2$ finite for terrestrial neutrino 
oscillations, and $\delta m^2$ finite and $m^2=\infty$ for solar neutrino
oscillations.  If  $\delta m^2$ is close to the best-fit to solar neutrino 
data ($\delta m^2\sim 5\times 10^{-6}$ eV$^2$)  {\em and\/} if $m^2$ is 
close to the  best-fit to atmospheric neutrino data 
($m^2\sim 0.6\times 10^{-2}$ eV$^2$) then  $\delta m^2/m^2\sim 10^{-3}$ 
and the zeroth order approximation is certainly adequate. In 
Ref.~\cite{Mo95} we have numerically shown that for $\delta m^2/m^2$
as high as $1/10$ the leading first-order corrections to the zeroth 
approximation  do not alter substantially the results of both the solar 
and the atmospheric neutrino data fit. However, if one takes the 
{\em highest\/} values of $\delta m^2$ allowed by solar neutrinos 
($\delta m^2\sim 1.5\times 10^{-4}$ eV$^2$) \cite{Mo96} and the 
{\em lowest\/} values of $m^2$ allowed by atmospheric neutrinos 
($m^2\sim 6\times 10^{-4}$ eV$^2$)  at $90\%$ C.L., then the ratio  
$\delta m^2/m^2$ is about 1/4, so that the two squared mass differences
are not well-separated, and our approximations become very rough. Such a 
contrived situation seems improbable, but if it were realized in nature, 
then one should necessarily  resort to the most general three-flavor 
formalism to analyze it.

%%%%%%%%%%%%%%%%%%%%%%%%%%%%%%%%%%%%%%%%%%%%%%%%%%%%%%%%%%%%%%%%%%%%%%%%%%%
\section{SUMMARY AND CONCLUSIONS}
%%%%%%%%%%%%%%%%%%%%%%%%%%%%%%%%%%%%%%%%%%%%%%%%%%%%%%%%%%%%%%%%%%%%%%%%%%%

	In this work we have analyzed the available  experimental data 
on the anomalous flavor composition of the atmospheric neutrino flux
with three-flavor neutrino oscillations. Data on upward-going muons are 
not included and will be examined in a separate work.

	The adopted theoretical framework  is characterized by one 
dominant neutrino square mass difference, $m^2$. The neutrino mass
spectrum can assume either form (a) or (b) of Fig.~1. Scenarios
(a) and (b) are physically different when neutrino oscillations in the 
earth matter background are considered. In both cases the variables 
relevant to atmospheric neutrino oscillations are $m^2$ and two mixing 
angles, $\psi$ and $\phi$.

	We have performed a global analysis of all data, and found the
regions of the $(m^2,\,\psi,\,\phi)$ parameter space in which
three-flavor neutrino oscillations are consistent with the  available 
data. In particular, we have included in the analysis the neutrino-induced
$e$-like and $\mu$-like event rates measured in four sub-GeV experiments,
Fr{\'e}jus, NUSEX, Kamiokande, and IMB, as well as the lepton rates 
measured in the five zenith-angle sectors of the multi-GeV Kamiokande 
experiment, for a total of 18 observables in the fit. We have made 
accurate calculations of the expected muon and electron rates in the 
various detectors, taking into account the differential energy-angle
distribution of the (anti)neutrino  fluxes, the differential
(anti)neutrino interaction cross-sections, and the detector efficiencies.
We have paid particular attention to the statistical analysis, which 
includes the proper correlations among the experimental and theoretical 
errors. The oscillation probabilities have been calculated in the 
three-flavor framework defined by Fig.~1, including the earth matter
effect in the evolution of the neutrino flavor states. The main results 
of our analysis are shown in Fig.~7 for scenario (a) and in Fig.~9 for 
scenario (b). Three-flavor neutrino oscillations provide a good fit to 
the 18 data (with three free parameter), as the minimum $\chi^2$ is 
$\sim 15$ in both cases.

	We have also analyzed in detail the subcases of two-flavor
$\nu_\mu\leftrightarrow\nu_\tau$ oscillations (Fig.~2) and
$\nu_\mu\leftrightarrow\nu_e$ oscillations (Fig.~4). The analysis of 
these subcases provided us with valuable information on the relative 
weight and influence of the different pieces of data  in the global fit, 
as well as on the importance of the earth matter effect (as derived by 
comparing Fig.~4 and Fig.~6). The Kamiokande multi-GeV data are fitted 
better in the $\nu_\mu\leftrightarrow\nu_e$ case than in the 
$\nu_\mu\leftrightarrow\nu_\tau$ case, as shown in Fig.~5, although 
more data are needed to confirm this indication. In particular, the 
upcoming data from the running SuperKamiokande experiment \cite{SKam}
will certainly help in clarifying the atmospheric neutrino anomaly and 
its implications in terms of neutrino properties.

	We have compared the region preferred by atmospheric neutrino 
data with the bounds coming from negative oscillation searches at 
accelerator and reactors \cite{Sc95} in Fig.~11. These bounds exclude
a large part  of the region allowed by atmospheric data  (especially 
in the limit of pure $\nu_\mu\leftrightarrow\nu_e$  oscillations) for
$m^2\gtrsim 10^{-2}$ eV$^2$. For $m^2\lesssim 10^{-2}$ eV$^2$ there 
are no significant bounds from reactor data and the atmospheric data 
fit is unaffected. In particular,  for $m^2\lesssim 10^{-2}$ eV$^2$ 
the threefold maximal mixing scenario \cite{Ki95,Ha95} is allowed by 
the data. It must be added, however, that threefold maximal mixing is 
not supported by the independent analysis of all solar neutrino data 
\cite{Mo96}.

	In Sec.~V~C we have also  discussed the interplay between 
atmospheric and solar neutrino results. Solar neutrino data place an 
upper bound on $\phi$  \cite{Mo96} that further constrains the 
atmospheric results at any $m^2$. The value of $m^2$ needed to fit 
atmospheric neutrino data is not compatible, within this framework,
with the possible recent indication for neutrino  oscillations coming 
from the LSND experiment \cite{LNS2}. A marginal compatibility between 
the LSND data and the atmospheric anomaly might be reached \cite{Ca96} 
if the information  coming from the zenith-angle distribution of
multi-GeV event is discarded.

	This work is part of a wider research programme in which we
intend to analyze, in  the same three-flavor framework, the world  data 
related to neutrino oscillations.  We have analyzed  so far the results 
coming from 14 experiments: 3 accelerator experiments (CERN CDHSW, 
Fermilab E531, and BNL E776) \cite{Sc95}, 3 reactor experiments 
(Bugey, G{\"o}sgen, and Krasnoyarsk reactors) \cite{Sc95}, 4 solar 
neutrino experiments (Homestake, GALLEX, SAGE, and Kamiokande) 
\cite {Mo96},  and 4 atmospheric neutrino experiments (Fr{\'e}jus, 
NUSEX, IMB, and Kamiokande, this work).  (See also \cite{Li94,Mo95}.)
We have discussed in \cite{Sc95} some implications of older LSND 
results \cite{LNS1}, and in \cite{Fo96} the tests of three-flavor mixing
in future long-baseline neutrino oscillation experiments. We hope that 
the three-flavor framework adopted in these works can become a popular 
way of analyzing or even presenting the experimental results or 
expectations, instead of the usual two-generation approach which is 
unable to accommodate more than one oscillation channel at a time.

%%%%%%%%%%%%%%%%%%%%%%%%%%%%%%%%%%%%%%%%%%%%%%%%%%%%%%%%%%%%%%%%%%%%%%%%%%%
\acknowledgements

	We thank  C.~Giunti and P.~I.~Krastev for useful discussions. 
We are grateful to J.~Pantaleone for  reading the manuscript and for 
very helpful comments. G.L.F.\ would like to thank Prof.~Milla Baldo 
Ceolin and the  organizers of the VII Workshop {\em Neutrino
Telescopes '96\/} (Venice, Italy, feb.\ 1996), where preliminary
results of this work were presented.  The work of E.L. was supported in 
part by INFN and in part by a Hansmann membership at the Institute for 
Advanced Study. This work was performed under the auspices of the 
European Theoretical Astroparticle Network (TAN).

%%%%%%%%%%%%%%%%%%%%%%%%%%%%%%%%%%%%%%%%%%%%%%%%%%%%%%%%%%%%%%%%%%%%%%%%%%%
\appendix
\section{Treatment of Kamiokande Multi-G{\lowercase{e}}V data}
\label{app:multigev}
%%%%%%%%%%%%%%%%%%%%%%%%%%%%%%%%%%%%%%%%%%%%%%%%%%%%%%%%%%%%%%%%%%%%%%%%%%%

	In this appendix we describe in detail our treatment of the 
Kamiokande multi-GeV data.

	The analysis of the Kamiokande multi-GeV data \cite{Fu94}
depends crucially upon the distribution of lepton events ($N_\beta$)
of given flavor $\beta$  as a function of the zenith angle $\theta$. In 
the presence of neutrino  oscillations, $\beta$-flavor lepton events may 
be initiated by  neutrinos of original flavor $\alpha$, and the angular 
distribution can be expressed as:
%..........................................................................
\begin{equation}\label{eq:dNdt}
\frac{dN_\beta}{d\theta}=\sum_\alpha\,
\int^\infty_{E_\nu^{\rm min}}\!\! dE_\nu\,
\frac{d^2\Phi_\alpha}{dE_\nu\,d\theta}\,
P_{\alpha\beta}\!
\int^{E_\nu}_{E_\ell^{\rm min}}\!\! dE_\ell\,
\frac{d\sigma_\beta}{dE_\ell}\,
\varepsilon_\beta\ ,
\end{equation}
%..........................................................................
where:
%..........................................................................
\begin{eqnarray}
\frac{dN_\beta(\theta)}{d\theta}                        &=&
{\rm lepton\ angular\ distribution}                         \ ,\nonumber\\
E_\nu                                                   &=&
{\rm neutrino\ energy}                                      \ ,\nonumber\\
E_\ell                                                  &=&
{\rm lepton\ energy}                                        \ ,\nonumber\\
\frac{d^2\Phi_\alpha(E_\nu,\,\theta)}{dE_\nu\,d\theta}  &=&
{\rm distribution\ of\ unoscillated\ }\nu_\alpha            \ ,         \\
P_{\alpha\beta}(E_\nu,\,\theta)                         &=&
{\rm flavor\ oscillation\ probability}                      \ ,\nonumber\\
\frac{d\sigma_\beta(E_\ell)}{dE_\ell}  &=&
{\rm differential\ }\nu_\beta{\rm\ cross\ section}          \ ,\nonumber\\
\varepsilon_\beta(E_\ell)                               &=&
{\rm lepton\ detection\ efficiency}                         \ .\nonumber
\end{eqnarray}
%..........................................................................

	For the sake of simplicity (and of computing time)  in 
Eq.~(\ref{eq:dNdt}) we have assumed that the lepton direction $\theta$
is the same as the incident neutrino direction $\theta_\nu$,
$\theta\simeq\theta_\nu$. Actually, in the Kamiokande multi-GeV data
sample, the typical difference is
$\sqrt{\langle(\theta-\theta_\nu)^2\rangle}\simeq15^{\circ}$--$20^{\circ}$
\cite{Fu94}. We simulate the effect of the  $\theta$-$\theta_\nu$ 
difference by smearing the distribution $dN_\beta/d\theta$ in 
Eq.~(\ref{eq:dNdt}) with a Gaussian distribution having a one-sigma
width of $\sim17^{\circ}$.

	The evaluation of the inner integral in  Eq.~(\ref{eq:dNdt}) 
requires detailed experimental information that is not available.
In particular, the  lepton detection efficiency function 
$\varepsilon_\beta(E_\ell)$,  which includes the intrinsic detector 
acceptance and the analysis cuts, is not published for multi-GeV 
data \cite{Fu94}.  A worse problem is due to the impossibility of
defining, event by event, the lepton energy $E_\ell$ for tracks that are 
not fully contained.  For these (partially contained) higher-energy events, 
the intrinsic ``total'' energy  can be associated with  the  released
``visible'' lepton energy on a statistical basis only, by means of a 
detailed simulation of the Kamiokande detector (which is beyond our 
possibilities and interests). However, the ignorance of such experimental 
ingredients can be overcome by re-writing Eq.~(\ref{eq:dNdt}) in terms of 
the simulated energy spectrum of the parent neutrinos (which embeds all 
these effects) as published  in \cite{Fu94}.

	The method is the following. The $(E_\nu,\,\theta)$-distribution 
of unoscillated neutrinos can be factorized as:
%..........................................................................
\begin{equation}\label{eq:factoriz}
\frac{d^2\Phi_\alpha(E_\nu,\,\theta)}{dE_\nu\,d\theta}=
\frac{d\Phi'_\alpha(E_\nu)}{dE\nu}\cdot
\frac{d\Phi''_\alpha(E_\nu,\,\theta)}{d\theta}\ ,
\end{equation}
%..........................................................................
with the normalization
%..........................................................................
\begin{equation}\label{eq:normaliz}
\int\!\! d\theta\, \frac{d\Phi''_\alpha(E_\nu,\,\theta)}{d\theta}=1 
\quad({\rm at\ any\ }E_\nu)\ .
\end{equation}
%..........................................................................

	Applying the factorization of Eq.~(\ref{eq:factoriz}) in 
Eq.~(\ref{eq:dNdt}) one has that:
%..........................................................................
\begin{eqnarray}\label{eq:newdNdt}
\frac{dN_\beta}{d\theta}
&=&\int\!\!dE_\nu\,\frac{d\Phi''_\alpha}{d\theta}\,P_{\alpha\beta}\,
\frac{d\Phi'_\alpha}{dE_\nu}\!\int\!\!dE_\ell\,
\frac{d\sigma_\beta}{dE_\ell}\,\varepsilon_\beta
                                                              \nonumber\\
&=&\int\!\!dE_\nu\,\frac{d\Phi''_\alpha}{d\theta}\,P_{\alpha\beta}\,
Q_{\alpha\beta}\,\frac{dn_\beta}{dE_\nu}\ ,
\end{eqnarray}
%..........................................................................
where
%..........................................................................
\begin{equation}
Q_{\alpha\beta}(E_\nu)=\frac{d\Phi'_\alpha}{dE_\nu}
\left(\frac{d\Phi'_\beta}{dE_\nu}\right)^{-1}
\end{equation}
%..........................................................................
is known from atmospheric flux calculations, and 
%..........................................................................
\begin{equation}
\frac{dn_\beta(E_\nu)}{dE_\nu}=\frac{d\Phi'_\beta}{dE_\nu}
\int\!\!dE_\ell\,\frac{d\sigma_\beta}{dE_\ell}\,
\varepsilon_\beta
\end{equation}
%..........................................................................
represents the energy distribution of the parent neutrinos that
induce $\beta$-lepton events in the Kamiokande detector, integrated
over the lepton spectrum.

	As reference energy-angle neutrino flux distributions, we use
the calculations of the Bartol group \cite{Ba89} smoothly connected to the 
Volkova calculations \cite{Vo80} at higher energies.  This reference 
choice corresponds to the option ``Flux~B'' in \cite{Fu94}.

	The function $dn_\beta(E_\nu)/dE_\nu$, that embeds all those 
experimental aspects of the lepton detection efficiency and energy 
reconstruction  that we ignore, is published in Fig.~2(b) of 
Ref.~\cite{Fu94}. Since both $d\Phi''_\alpha/{d\theta}$ and 
$Q_{\alpha\beta}$ are known, and $P_{\alpha\beta}$ is calculable in a 
given oscillation scenario, one finally has all the ingredients to 
calculate the angular distribution of lepton events from 
Eq.~(\ref{eq:newdNdt}).%
%---------------------
\footnote{	It should be added, however, that in this way our analysis
		of the multi-GeV data depends implicitly upon the neutrino 
		cross sections as implemented in the Kamiokande Monte~Carlo 
		simulation.}
%---------------------

	In particular, we have computed the angular distribution of
$e$-like and $\mu$-like multi-GeV events in absence of oscillations
($P_{\alpha\beta}=\delta_{\alpha\beta}$). The results are shown in Fig.~12
as dashed lines. The solid lines represent the published Kamiokande
simulation  \cite{Fu94}. The agreement is very good, as the differences 
are smaller than the statistical uncertainties.

	The author of Ref.~\cite{Ya96}  used independently a somewhat 
similar approach to the analysis of multi-GeV data but did not obtain, 
however,  a good agreement with the published Kamiokande simulation of
$e$-like events.

	A final remark is in order. The functions 
$dn_\beta(E_\nu)/dE_\nu$ $(\beta=e,\,\mu)$ reported in  \cite{Fu94}
include the contributions of both neutrinos and antineutrinos. However, it
 is important to separate the $\nu$ and $\bar\nu$ contributions, since
$P(\nu_\alpha\to\nu_\beta)\neq P(\bar\nu_\alpha\to\bar\nu_\beta)$ when 
matter oscillations  are considered. We make the reasonable assumption that 
$\varepsilon_\beta\simeq\varepsilon_{\bar\beta}$. Then, since the ratios
of $\nu$ and $\bar\nu$ fluxes and cross sections are known at any energy,
one can separate the distribution $dn_\beta/dE_\nu$ of parent 
{\em neutrinos\/} from the distribution $dn_{\bar\beta}/dE_\nu$
of parent  {\em antineutrinos\/}.

%%%%%%%%%%%%%%%%%%%%%%%%%%%%%%%%%%%%%%%%%%%%%%%%%%%%%%%%%%%%%%%%%%%%%%%%%%%
\section{Correlation of errors}
\label{app:correlations}
%%%%%%%%%%%%%%%%%%%%%%%%%%%%%%%%%%%%%%%%%%%%%%%%%%%%%%%%%%%%%%%%%%%%%%%%%%%

	In \cite{Li95} we  examined the various sources of uncertainties
affecting the measured and expected $e$-like and $\mu$-like  event rates
$R_e$ and $R_\mu$ in the atmospheric neutrino experiments, together
with their correlations. We were thus able, for any single experiment,
to construct the covariance matrix of the residuals 
($R_\alpha^{\rm theor}-R_\alpha^{\rm expt}$) and to perform a correct 
(Gaussian) statistical analysis of the atmospheric $\nu$ anomaly. As far 
as a single experiment is concerned, here we use the same approach.

	However, when the information of two or more experiments is 
combined---as in  the present work---one has also to take into account 
that the theoretical errors of the neutrino fluxes are highly correlated 
from experiment to experiment. For instance, a hypothetical systematic 
shift of $+20\%$ in the calculated (unoscillated)  flux of sub-GeV 
$\nu_e$'s  propagates coherently to the expected rates
of $e$-like events in {\em all\/} the sub-GeV experiments at the 
same time. Moreover, one also expects the $\mu$-like event rates to 
increase by $\sim20\%$ because of the tight correlation of calculated
$\nu_e$ and $\nu_\mu$ fluxes, with an allowance for a residual 
uncertainty (of about $5\%$) in the $\mu/e$ ratio.

	Therefore, in constructing the covariance matrix for the
observables analyzed in this work, we include the  additional off-diagonal 
elements corresponding to the correlations of the neutrino flux 
uncertainties between any two experiments, and between any two bins of 
the Kamiokande multi-GeV data sample.%
%-----------------------------------------------
\footnote{	In principle there could also be correlations among the 
		{\em experimental\/}  systematic uncertainties
		affecting any two bins of the Kamiokande multi-GeV data. 
		However, for lack of published information 
		\protect\cite{Fu94}  we ignore such additional 
		correlations.}
%----------------------------------------------

	More precisely, let us call $(A,\,B)$ a generic couple of 
experiments (or couple of multi-GeV data bins). Then the correlations 
between the $\nu_e$ and  $\nu_\mu$ theoretical flux errors in $A$ and $B$ 
are given by:
%..........................................................................
\begin{eqnarray}\label{eq:corr}
\rho_{ee}     (A,\,B) &=& 1\ ,\nonumber\\
\rho_{\mu\mu} (A,\,B) &=& 1\ , \\
\rho_{\mu e}  (A,\,B) &=& 1-\frac{1}{2}
\frac{\sigma^2_{\rm flux}}{\sigma^2_{\rm ratio}}
\ ,\nonumber
\end{eqnarray}
%..........................................................................
where $\sigma_{\rm flux}$ is the  assumed fractional uncertainty in the 
overall flux normalization (e.g., $\pm 30\%$), and $\sigma_{\rm ratio}$ 
is the assumed  residual uncertainty in the $\mu/e$ ratio (typically 
$\pm 5\%$, which we choose as default value). For instance, for 
$(\sigma_{\rm flux},\,\sigma_{\rm ratio})=(30\%,\,5\%)$ the $\mu$-$e$ 
flux error correlation is $\rho_{\mu e}=0.986$ \cite{Li95}.

	We have decided, however, to ignore the correlations when A 
labels a sub-GeV observable {\em and\/} B labels a multi-GeV observable 
(or vice~versa). In fact, the flux of low-energy and high-energy neutrinos 
are not necessarily correlated. A systematic shift of, e.g., $+20\%$ in 
the low-energy  neutrino flux normalization does not necessarily imply the 
same shift at higher energy. This would happen, for instance, if the 
{\em slope\/} of the theoretical neutrino energy distribution were 
systematically biased.  At present, we do not know how to relate the 
uncertainties affecting the low-energy and the high-energy fluxes of 
atmospheric neutrinos, and thus ignore their possible correlations
in the $\chi^2$ statistics.

	The inclusion of the (known) correlation effects in any single 
experiment, as well as in the combination of all the experimental data, 
is a distinguishing feature of our analysis.

%%%%%%%%%%%%%%%%%%%%%%%%%%%%%%%%%%%%%%%%%%%%%%%%%%%%%%%%%%%%%%%%%%%%%%%%%%%
\section{Symmetries  of the oscillation probability}
\label{app:symmetries}
%%%%%%%%%%%%%%%%%%%%%%%%%%%%%%%%%%%%%%%%%%%%%%%%%%%%%%%%%%%%%%%%%%%%%%%%%%%

	In this appendix we discuss several symmetry properties of the 
neutrino and antineutrino oscillation probabilities under given 
transformations of the neutrino masses and mixing, in the two scenarios 
(a) and (b) of Fig.~1. These properties are useful to understand the 
results of the analysis of the atmospheric neutrino data. In particular, 
we show that the scenarios (a) and (b) are not equivalent when the earth 
matter effects are included in the (anti)neutrino propagation.

	We recall that we always assume $m^2$ positive $(m^2=|m^2|)$, and 
that the two scenarios (a) and (b) are distinguished by the overall sign of 
$m^2$: $({\rm a})\to({\rm b})\Longleftrightarrow +m^2\to-m^2$. It is useful 
to set conventionally the zero of the neutrino squared mass scale halfway 
between the doublet $(\nu_1,\,\nu_2)$ and the ``lone'' state $\nu_3$ 
shown in Fig.~1. With this position,  the neutrino squared mass spectrum 
takes the form:
%..........................................................................
\begin{equation}\label{eq:scenarios}
(m^2_1,\,m^2_2,\,m^2_3)=\left\{\begin{array}{cl}
(-\frac{m^2}{2},\,-\frac{m^2}{2},\,+\frac{m^2}{2}) & {\rm\ \ scenario\ (a)}\\
(+\frac{m^2}{2},\,+\frac{m^2}{2},\,-\frac{m^2}{2}) & {\rm\ \ scenario\ (b)}
\end{array}\right. \ .
\end{equation}
%..........................................................................

	Let us consider the following 8 transformations $T_i$:
%..........................................................................
\begin{equation}\label{eq:trans}
\begin{array}{ccrclcl}
T_1: & \quad &  m^2         & \rightarrow & -m^2                & \quad &
                                             {\rm at\ any\ }\psi,\,\phi\ ;\\ 
T_2: & \quad & (m^2,\,\phi) & \rightarrow & (m^2,\,\pi/2-\phi)  & \quad &
                                                       {\rm at\ }\psi=0\ ;\\
T_3: & \quad & (m^2,\,\phi) & \rightarrow & (m^2,\,\pi/2-\phi)  & \quad &
                                                   {\rm at\ }\psi=\pi/2\ ;\\
T_4: & \quad & (m^2,\,\psi) & \rightarrow & (m^2,\,\pi/2-\psi)  & \quad &
                                                       {\rm at\ }\phi=0\ ;\\
T_5: & \quad & (m^2,\,\phi) & \rightarrow & (-m^2,\,\pi/2-\phi) & \quad &
                                                       {\rm at\ }\psi=0\ ;\\
T_6: & \quad & (m^2,\,\phi) & \rightarrow & (-m^2,\,\pi/2-\phi) & \quad &
                                                   {\rm at\ }\psi=\pi/2\ ;\\
T_7: & \quad & (m^2,\,\psi) & \rightarrow & (-m^2,\,\pi/2-\psi) & \quad &
                                                       {\rm at\ }\phi=0\ ;\\
T_8: & \quad & (\nu,\,m^2)  & \rightarrow & (\bar\nu,\,-m^2)    & \quad &
                                           {\rm at\ any\ }\psi,\,\phi\ .  \\ 
\end{array}
\end{equation}
%..........................................................................

	The transformation $T_1$ changes the overall sign of $m^2$ and thus
maps scenario (a) into (b) or vice versa. The transformations $T_{2,5}$,  
$T_{3,6}$, and  $T_{4,7}$  are relevant respectively for the subcases of 
pure two-flavor $\nu_e\leftrightarrow\nu_\tau$ oscillations $(\psi=0)$,
$\nu_e\leftrightarrow\nu_\mu$ oscillations $(\psi=\pi/2)$, and
$\nu_\mu\leftrightarrow\nu_\tau$ oscillations $(\phi=0)$. The 
transformation $T_8$ interchanges neutrinos with antineutrinos and, 
at the same time,  changes the sign of $m^2$. Notice that the $T_i$'s in 
Eq.~(\ref{eq:trans}) are not all independent:
$T_5=T_1\cdot T_2$,
$T_6=T_1\cdot T_3$, and
$T_7=T_1\cdot T_4$.

	We prove the following statements: 
(1) 	in vacuum, the oscillation probabilities are invariant under 
	$T_1$, $T_2$, $T_3$, $T_4$, $T_5$, $T_6$, $T_7$, and $T_8$;
(2) 	in matter, the oscillation probabilities are invariant only
	under $T_4$, $T_5$, $T_6$, $T_7$, and $T_8$;
(3) 	in matter, the additional symmetries $T_1$, $T_2$, and $T_3$
	are restored in the limit $m^2\to\infty$.

	The statement (1) is evident from inspection of the vacuum
oscillation probabilities (equal for neutrinos and antineutrinos):
%..........................................................................
\begin{eqnarray}\label{eq:Pvac}
P_{ee}^{\rm vac}       & = & 1-4s^2_\phi c^2_\phi\,S\ ,
\nonumber\\
P_{\mu\mu}^{\rm vac}   & = & 1-4c^2_\phi s^2_\psi(1-c^2_\phi s^2_\psi)\,S\ , 
\nonumber\\
P_{\tau\tau}^{\rm vac} & = & 1-4c^2_\phi c^2_\psi(1-c^2_\phi c^2_\psi)\,S\ ,
         \\
P_{e\mu}^{\rm vac}     & = & 4s^2_\phi c^2_\phi s^2_\psi\,S\ ,             
\nonumber\\
P_{e\tau}^{\rm vac}    & = & 4s^2_\phi c^2_\phi c^2_\psi\,S\ ,   
\nonumber\\
P_{\mu\tau}^{\rm vac}  & = & 4c^4_\phi s^2_\psi c^2_\psi\,S\ , 
\nonumber
\end{eqnarray}
%..........................................................................
where $S=\sin^2(m^2 x/4E_\nu)$. Note that the angles $\omega$ and 
$\delta$ never appear in  Eq.~(\ref{eq:Pvac}). Of course, 
$P_{\alpha\beta}=P_{\beta\alpha}$.

	The probabilities in Eq.~(\ref{eq:Pvac}) are invariant under 
$T_1$, implying that the cases (a) and (b) are indistinguishable in 
vacuum.  The symmetries $T_2$ and $T_5$ imply that the parameters of pure
$\nu_e\leftrightarrow\nu_\tau$ oscillations in vacuum can be restricted to
the case $+m^2$ and $\phi\in[0,\,\pi/4]$, as the cases $-m^2$  
$\phi\in[\pi/4,\,\pi/2]$ become equivalent. Analogously, this is true 
for pure $\nu_e\leftrightarrow\nu_\mu$ or 
$\nu_\mu\leftrightarrow\nu_\tau$ oscillations in vacuum.

	When matter effects are included, the situation becomes more 
complicated \cite{Pa94} and several symmetries are broken.  For the sake 
of simplicity, we discuss the symmetry properties of the oscillation 
probabilities in the  case of constant electron density, $N_e={\rm const}$. 
Our conclusions are also valid  for a generic $N_e=N_e(x)$, but the proof is
considerably more involved and less transparent,  since the neutrino 
propagation equations are not analitically integrable for a generic density. 
Instead, for  constant $N_e$  the  oscillation  probabilities can be 
expressed in compact form. For {\em neutrinos\/} they are given by 
(we omit the derivation):
%..........................................................................
\begin{eqnarray}\label{eq:Pmat}
P_{ee}       &=& 1-4s^2_\Phi c^2_\Phi         \,S_{31}\ ,\nonumber\\
P_{\mu\mu}   &=& 1-4s^2_\Phi c^2_\Phi s^4_\psi\,S_{31}
                  -4s^2_\Phi s^2_\psi c^2_\psi\,S_{21}
                  -4c^2_\Phi s^2_\psi c^2_\psi\,S_{32}\ ,\nonumber\\
P_{\tau\tau} &=& 1-4s^2_\Phi c^2_\Phi c^4_\psi\,S_{31}
                  -4s^2_\Phi s^2_\psi c^2_\psi\,S_{21}
                  -4c^2_\Phi s^2_\psi c^2_\psi\,S_{32}\ ,         \\
P_{e\mu}     &=&   4s^2_\Phi c^2_\Phi s^2_\psi\,S_{31}\ ,\nonumber\\
P_{e\tau}    &=&   4s^2_\Phi c^2_\Phi c^2_\psi\,S_{31}\ ,\nonumber\\
P_{\mu\tau}  &=&  -4s^2_\Phi c^2_\Phi s^2_\psi c^2_\psi\,S_{31}
                  +4s^2_\Phi s^2_\psi c^2_\psi\,S_{21}
                  +4c^2_\Phi s^2_\psi c^2_\psi\,S_{32}\ ,\nonumber
\end{eqnarray}
%..........................................................................
where $\Phi$ is the effective mixing angle $\phi$ in matter ($\psi$
remains unchanged):
%..........................................................................
\begin{equation}\label{eq:Phi}
\sin 2\Phi = \frac{\sin2\phi}%
{\sqrt{(\cos2\phi\mp A/m^2)^2+(\sin2\phi)^2}}\ .
\end{equation}
%..........................................................................
In Eq.~(\ref{eq:Pmat}) the oscillating factors $S_{ij}$ are defined as
%..........................................................................
\begin{equation}\label{eq:S}
S_{ij}=\sin^2\left( \frac{M^2_i-M^2_j}{4E_\nu}x\right)\ ,
\end{equation}
%..........................................................................
where $M^2_i$ are the neutrino square mass eigenvalues in matter,
%..........................................................................
\begin{eqnarray}\label{eq:M}
M^2_1 &=&\mp\frac{m^2}{2}\frac{s_{2\phi}}{s_{2\Phi}}+\frac{A}{2}\ ,\nonumber\\
M^2_2 &=&\mp\frac{m^2}{2}                                               \ , \\
M^2_3 &=&\pm\frac{m^2}{2}\frac{s_{2\phi}}{s_{2\Phi}}+\frac{A}{2}\ ,\nonumber
\end{eqnarray}
%..........................................................................
and $A=2\sqrt{2}G_F N_e E_\nu$ is the matter-induced square mass term.
We recall that the neutrino square mass eigenvalues in vacuum
are given in Eq.~(\ref{eq:scenarios}).

	In Eqs.~(\ref{eq:Phi})--(\ref{eq:M}) the upper sign refer to scenario
(a) and the lower sign to scenario (b). Notice that, as in the vacuum case,
 the angles $\omega$ and $\delta$ never appear in the oscillation 
probabilities in  Eq.~(\ref{eq:Pmat}).

	From Eqs.~(\ref{eq:Pmat})--(\ref{eq:M}) it follows that the earth 
matter effects do not vanish for threefold maximal mixing, corresponding to
$(\tan^2\psi,\,\tan^2\phi)=(1,\,1/2)$. In the interesting range 
$10^{-3}{\rm\ eV}^2\lesssim m^2 \lesssim 10^{-2}{\rm\ eV}^2$, the matter 
corrections to the ``vacuum'' lepton rates  can be as large as $10\%$ in 
the sub-GeV case and as large as $30\%$ in the multi-GeV case. The fits 
to the threefold maximal mixing scenario in \cite{Ki95,Ha95} were performed 
neglecting matter effects.

	The oscillation probabilities in Eq.~(\ref{eq:Pmat}) are not 
invariant under the transformation $T_1$. Therefore, the scenarios (a) and 
(b) are  in general physically different for atmospheric neutrinos 
traversing the earth matter. The difference can be traced to the 
matter-induced neutrino mass term $A$ [Eq.~(\ref{eq:M})], which is
positive both in (a) and (b), while the overall sign of the vacuum mass 
$m^2$ changes in the two scenarios.

	However, the probabilities in Eq.~(\ref{eq:Pmat}) are invariant 
under the transformations $T_4$, $T_5$, $T_6$, and $T_7$. Notice in 
particular that:
%..........................................................................
\begin{equation}
\begin{array}{lccccl}
\psi=0     & (\nu_e\leftrightarrow\nu_\tau) &  & \rightarrow &  &
P_{e\tau}=4s^2_\Phi c^2_\Phi   S'\ ,          \\
\psi=\pi/2 & (\nu_e\leftrightarrow\nu_\mu)  &  & \rightarrow &  &
P_{e\mu}=4s^2_\Phi c^2_\Phi    S'\ ,          \\
\phi=0     &(\nu_\mu\leftrightarrow\nu_\tau)&  & \rightarrow &  &
P_{\mu\tau}=4s^2_\psi c^2_\psi S = P_{\mu\tau}^{\rm vac}\ ,
\end{array}
\end{equation}
%..........................................................................
where
%..........................................................................
\begin{equation}
S'=
\sin^2\left(\frac{m^2}{4E_\nu}\frac{s_{2\phi}}{s_{2\Phi}}x \right)\ .
\end{equation}
%..........................................................................

	The symmetry $T_3$ does not apply to the probabilities in
Eq.~(\ref{eq:Pmat}). Therefore, the description of pure two-flavor 
oscillation $\nu_e\leftrightarrow\nu_\mu$ in matter cannot be exhausted by
taking $+m^2$ and $\phi\in[0,\,\pi/4]$. Either  one takes $+m^2$ and 
extends the range of $\phi$  to $[0,\,\pi/2]$ (as in this work), or one 
keeps $\phi\in[0,\,\pi/4]$, but considers both $+m^2$  and $-m^2$ (as in 
\cite{Ak93}). Analogously, this is true for $\nu_e\leftrightarrow\nu_\tau$ 
oscillations in matter, which are not invariant under $T_2$.

	Notice that the symmetry $T_4$, corresponding to 
$\psi\to\frac{\pi}{2}-\psi$  at given $m^2$ for pure 
$\nu_\mu\leftrightarrow\nu_\tau$ oscillations $(\phi=0)$, holds both in 
vacuum and in matter. In fact, matter effects are irrelevant for pure  
$\nu_\mu\leftrightarrow\nu_\tau$ oscillations.

	Let us now consider the {\em antineutrino\/} oscillation
probabilities. Equation~(\ref{eq:Pvac}) holds both for neutrinos and 
antineutrino oscillations (in vacuum). Equation~(\ref{eq:Pmat}) refers 
only to neutrinos. For antineutrinos, the  matter-induced term $A$ changes 
sign. However, if one also changes the sign of $m^2$, the antineutrino 
propagation becomes equivalent to neutrino propagation. In other words, 
the oscillation probabilities of {\em neutrinos\/} in scenario 
(a) are equal to the oscillation probabilities of {\em antineutrinos\/} 
in scenario (b), and vice~versa, as expressed by the symmetry $T_8$. 
Notice that the higher symmetry $T_1\cdot T_8$, which corresponds to 
$\nu\to\bar\nu$ at any given $(m^2,\,\psi,\,\phi)$, holds in vacuum only.

	It is amusing to notice that, in the purely hypothetical situation 
of equal $\nu$ and $\bar\nu$ fluxes at any energy $E_\nu$, and of equal
$\nu$ and $\bar\nu$ absorption cross sections, the symmetry $T_8$ would 
imply the same physics in atmospheric neutrino experiments in both scenarios 
(a) and (b).

	Finally, we consider the case of $m^2\to\infty$ or, more precisely,
$m^2\gg A$. In this limit, the oscillation probabilities in 
Eq.~(\ref{eq:Pmat}) become:
%..........................................................................
\begin{eqnarray}\label{eq:m2large}
P_{ee}       &=& P_{ee}^{\rm vac}                \ , \nonumber\\
P_{\mu\mu}   &=& P_{\mu\mu}^{\rm vac}-\delta P   \ , \nonumber\\
P_{\tau\tau} &=& P_{\tau\tau}^{\rm vac}-\delta P \ , \\
P_{e\mu}     &=& P_{\mu e}^{\rm vac}             \ , \nonumber\\
P_{e\tau}    &=& P_{e\tau}^{\rm vac}             \ , \nonumber\\
P_{\mu\tau}  &=& P_{\tau\mu}^{\rm vac}+\delta P  \ , \nonumber
\end{eqnarray}
%..........................................................................
where $\delta P = 4s^2_\phi s^2_\psi c^2_\psi\sin^2(A c^2_\phi x/4E_\nu)$.

	Notice  that in the limit $m^2\gg A$ not all the probabilities
in Eq.~(\ref{eq:m2large}) tend to their  vacuum value.  However, in this 
limit all the symmetries ($T_1,\,T_2,\,\dots,\,  T_8$) of the vacuum 
oscillation case apply. In the subcases of pure two-flavor oscillations 
($\psi=0$ or $\psi=\pi/2$ or $\phi=0$) one has $\delta P=0$ and 
Eq.~(\ref{eq:m2large}) simply reads  
$P_{\alpha\beta}=P_{\alpha\beta}^{\rm vac}$ (averaged vacuum oscillations
regime).

%%%%%%%%%%%%%%%%%%%%%%%%%%%%%%%%%%%%%%%%%%%%%%%%%%%%%%%%%%%%%%%%%%%%%%%%%%%

%%%%%%%%%%%%%%%%%%%%%%%%%%%%%%%%%%%%%%%%%%%%%%%%%%%%%%%%%%%%%%%%%%%%%%%%%%%%%%
%%		F I G U R E     C A P T I O N S
%%%%%%%%%%%%%%%%%%%%%%%%%%%%%%%%%%%%%%%%%%%%%%%%%%%%%%%%%%%%%%%%%%%%%%%%%%%%%%
%...........................................................................01
\begin{figure}
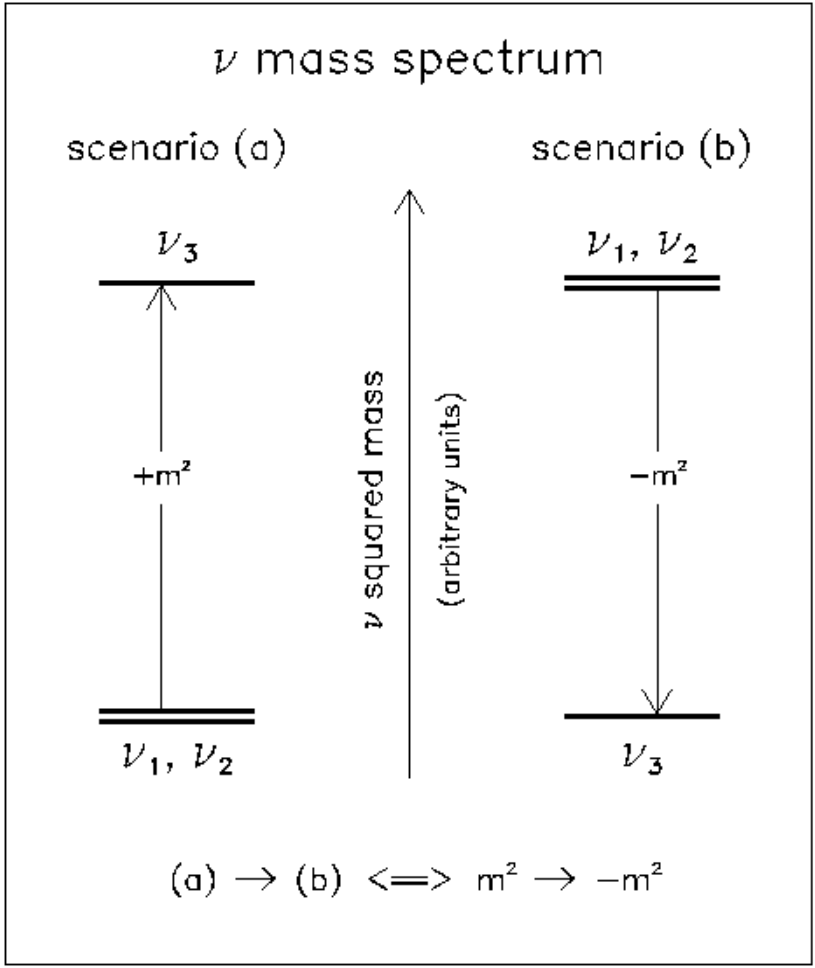

\caption{	The neutrino mass spectrum adopted in this work. One 
		neutrino mass eigenstate $(\nu_3)$ is assumed to be 
		largely separated from the quasi-degenerate doublet 
		$(\nu_1,\,\nu_2)$ by a square mass difference
		$|m^2_3-m^2_{1,2}|\simeq m^2$.  The two possible
		scenarios (a) and (b) are physically different when
		the earth matter effects are included in the atmospheric 
		$\nu$ propagation. The two square mass spectra in (a) and (b) 
		are related by: 
		${\rm (a)}\to{\rm (b)}\Longleftrightarrow m^2\to-m^2$.}
\end{figure}
%...........................................................................02
\begin{figure}
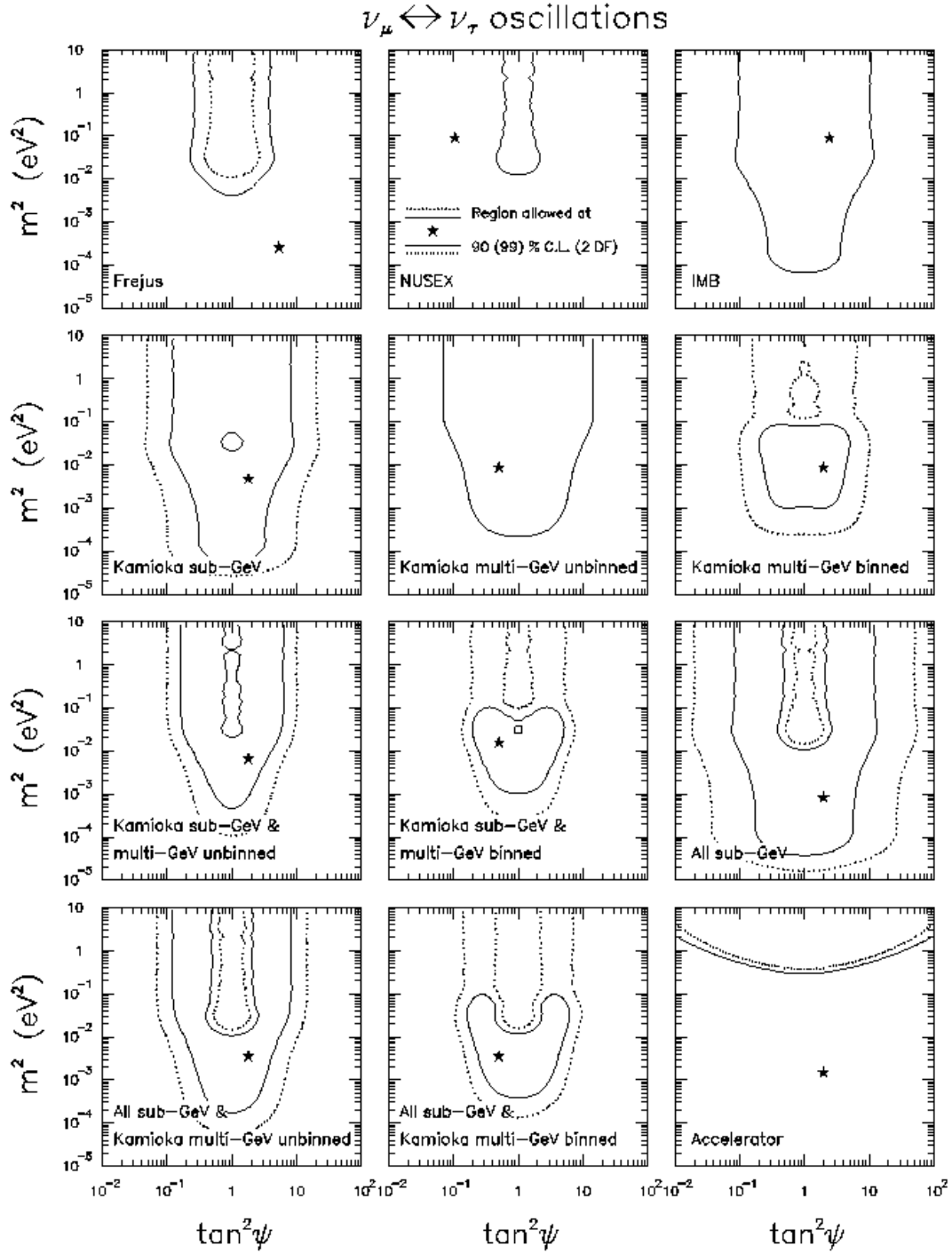

\caption{	Analysis of separate and combined atmospheric neutrino
		data assuming pure $\nu_\mu\leftrightarrow\nu_\tau$
		oscillations ($\phi=0$ in our framework) in the mass-mixing
		plane $(\tan^2\psi,\,m^2)$.  Contours at $90\%$ and $99\%$
		C.L.\ for $N_{\rm DF}=2$ are shown. The allowed regions
		are marked by stars. The last panel shows the constraints 
		coming from the combination of established accelerator 
		data. Notice the symmetry of all the allowed regions with 
		respect to the axis $\psi=\pi/4$.}
\end{figure}
%...........................................................................03
\begin{figure}
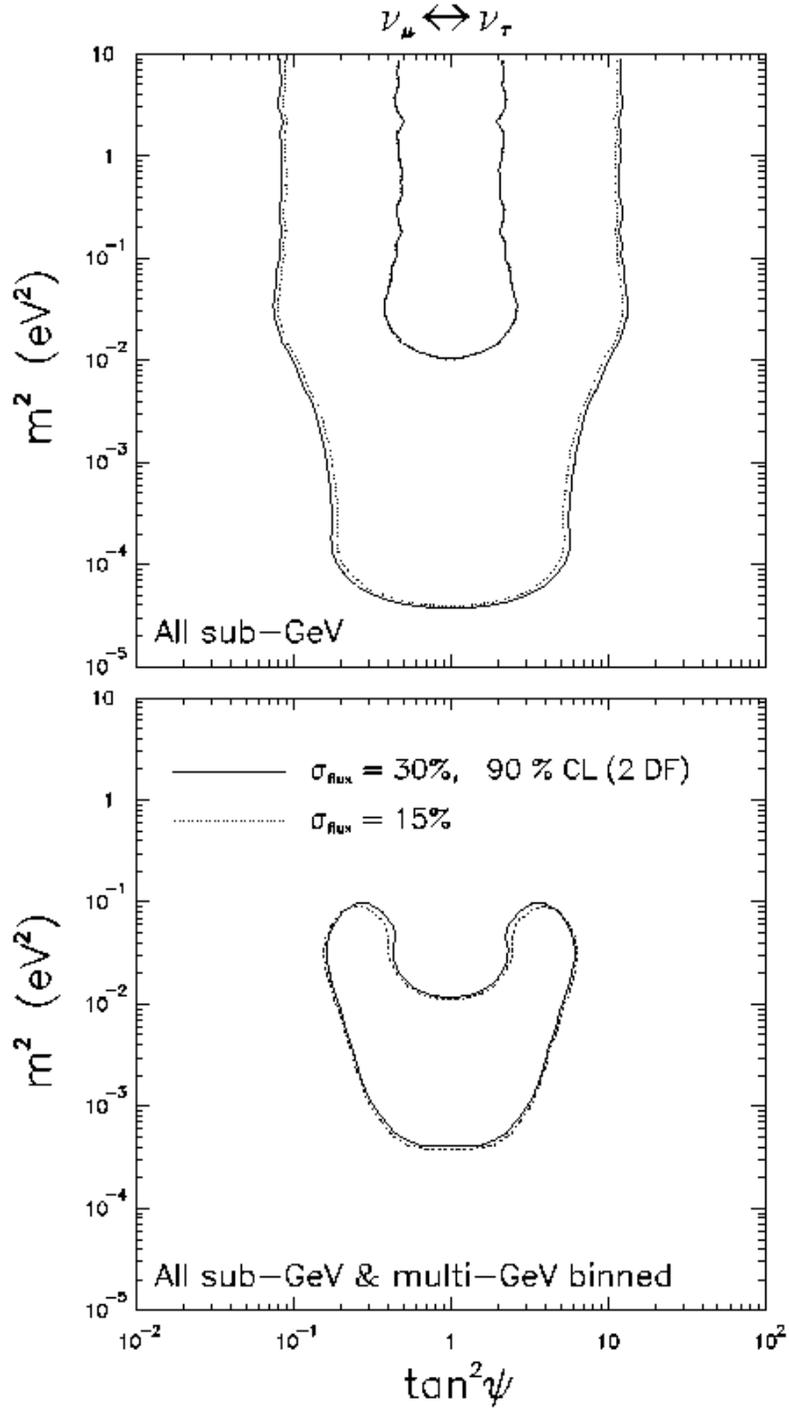

\caption{	Effects of the reduction of the neutrino flux uncertainty 
		from the default value of $\pm 30\%$ (solid lines) to 
		$\pm 15\%$ (dotted lines). The variations in the regions 
		allowed at $90\%$ C.L.\ by sub-GeV data (upper panel) 
		or sub-GeV+multi-GeV data (lower panel) are very small.}
\end{figure}
%...........................................................................04
\begin{figure}
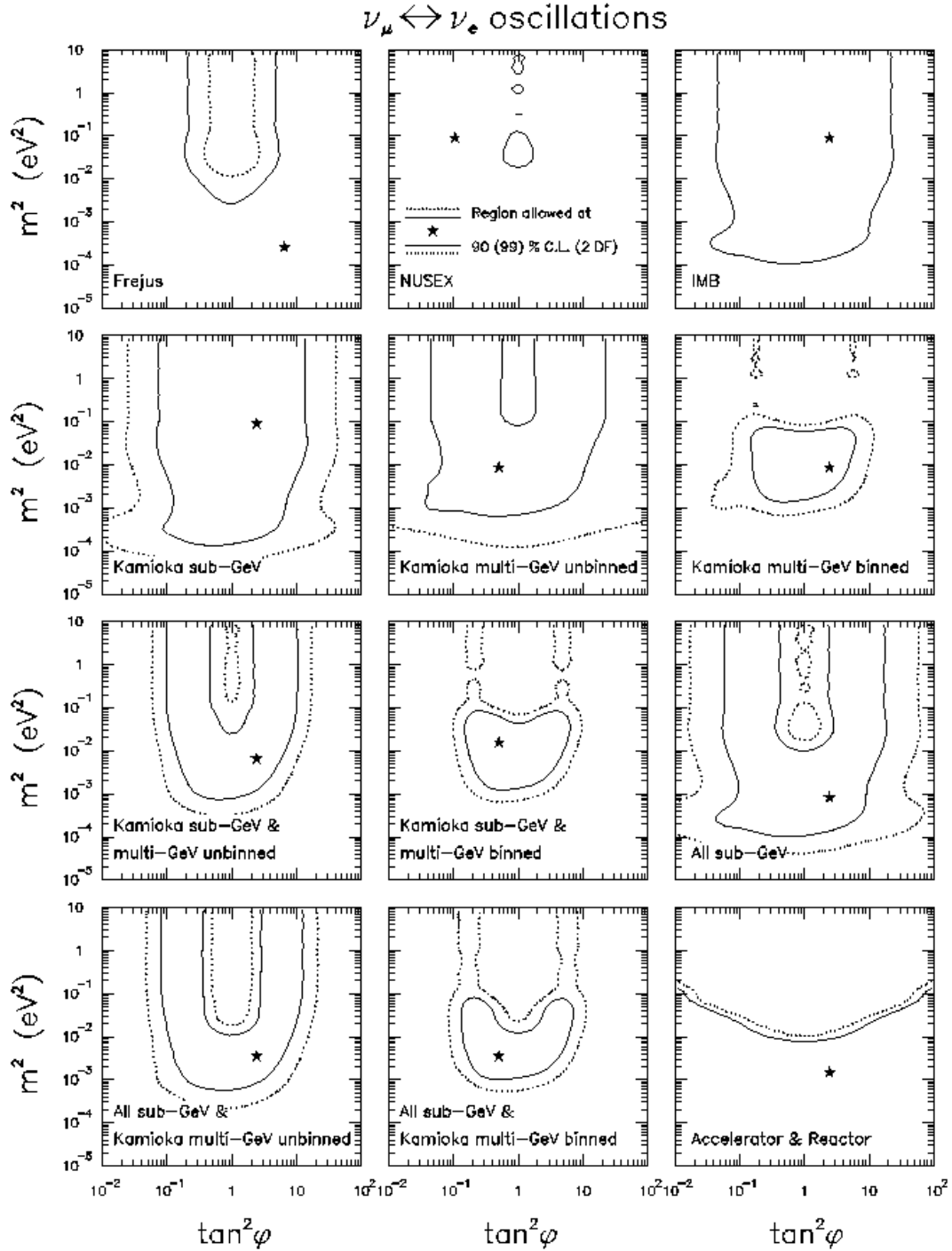

\caption{	Separate and combined analysis of atmospheric neutrino
		data assuming pure $\nu_\mu\leftrightarrow\nu_e$
		oscillations ($\psi=\pi/2$ in our framework), in the 
		mass-mixing plane $(\tan^2\phi,\,m^2)$. Contours at 
		$90\%$ (solid) and $99\%$ C.L.\ (dotted) for 
		$N_{\rm DF}=2$ are shown. The allowed regions are marked 
		by stars. The last panel shows the constraints coming 
		from the combination of established accelerator and 
		reactor data. The contours of the atmospheric $\nu$ 
		allowed regions are not symmetric with respect to the 
		axis $\phi=\pi/4$, due to matter oscillation effects.}
\end{figure}
%...........................................................................05
\begin{figure}
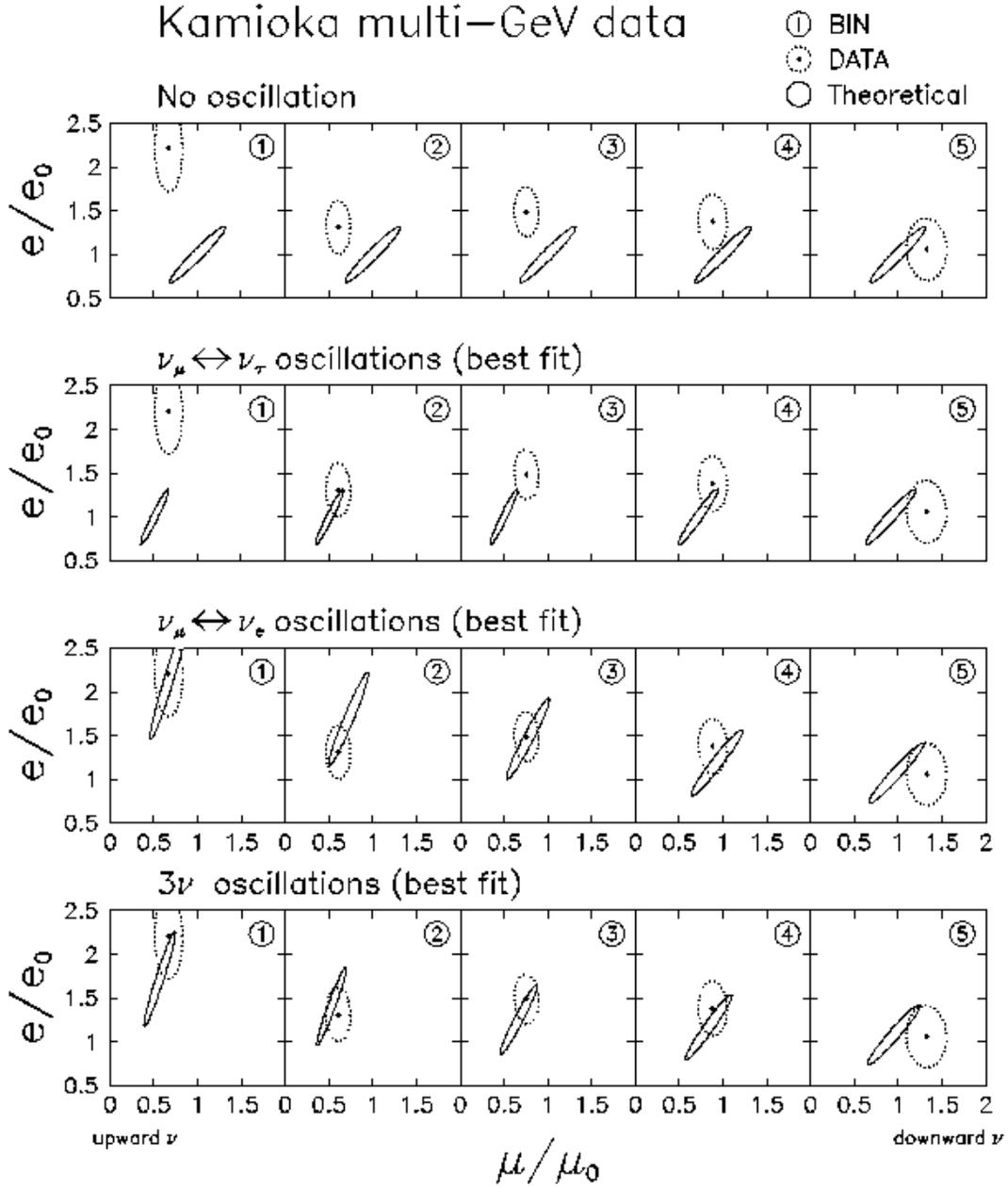

\caption{	Bin-by-bin analysis of multi-GeV Kamiokande data in
		the plane of the $\mu$ and $e$ lepton rates, normalized
		to their theoretical values without oscillations,
		$\mu_0$ and $e_0$. Solid ellipses: theoretical predictions 
		at $1\sigma$ level $(\Delta \chi^2=1)$. Dotted ellipses: 
		experimental data at $1\sigma$ level. Notice how the 
		theoretical ellipses change from the upper panel 
		(no oscillation) to the  three lower panels (best-fit 
		cases for two-flavor and three-flavor oscillations). The 
		fit is better in the $\nu_\mu\leftrightarrow\nu_e$ case 
		than in the $\nu_\mu\leftrightarrow\nu_\tau$ case. The
		overall fit improves slightly in the $3\nu$ oscillation case.}
\end{figure}
%...........................................................................06
\begin{figure}
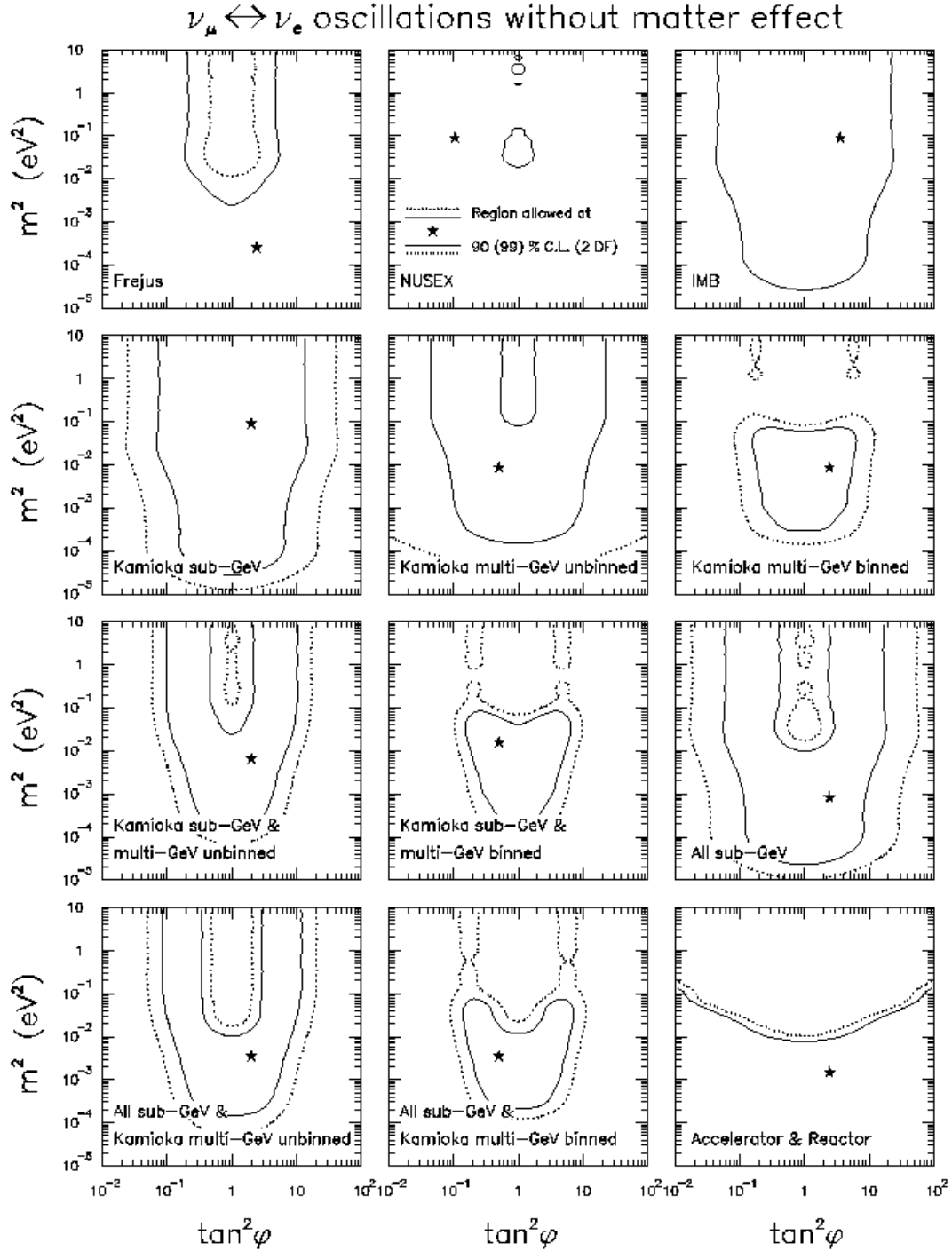

\caption{	As in Fig.~4, but excluding the earth matter effect
		(pure vacuum oscillations). All the contours are now 
		symmetric with respect to the axis $\phi=\pi/4$. The 
		regions allowed by the atmospheric neutrino data in the
		lower half of each panel are substantially different from 
		those reported in Fig.~4.}
\end{figure}
%...........................................................................07
\begin{figure}
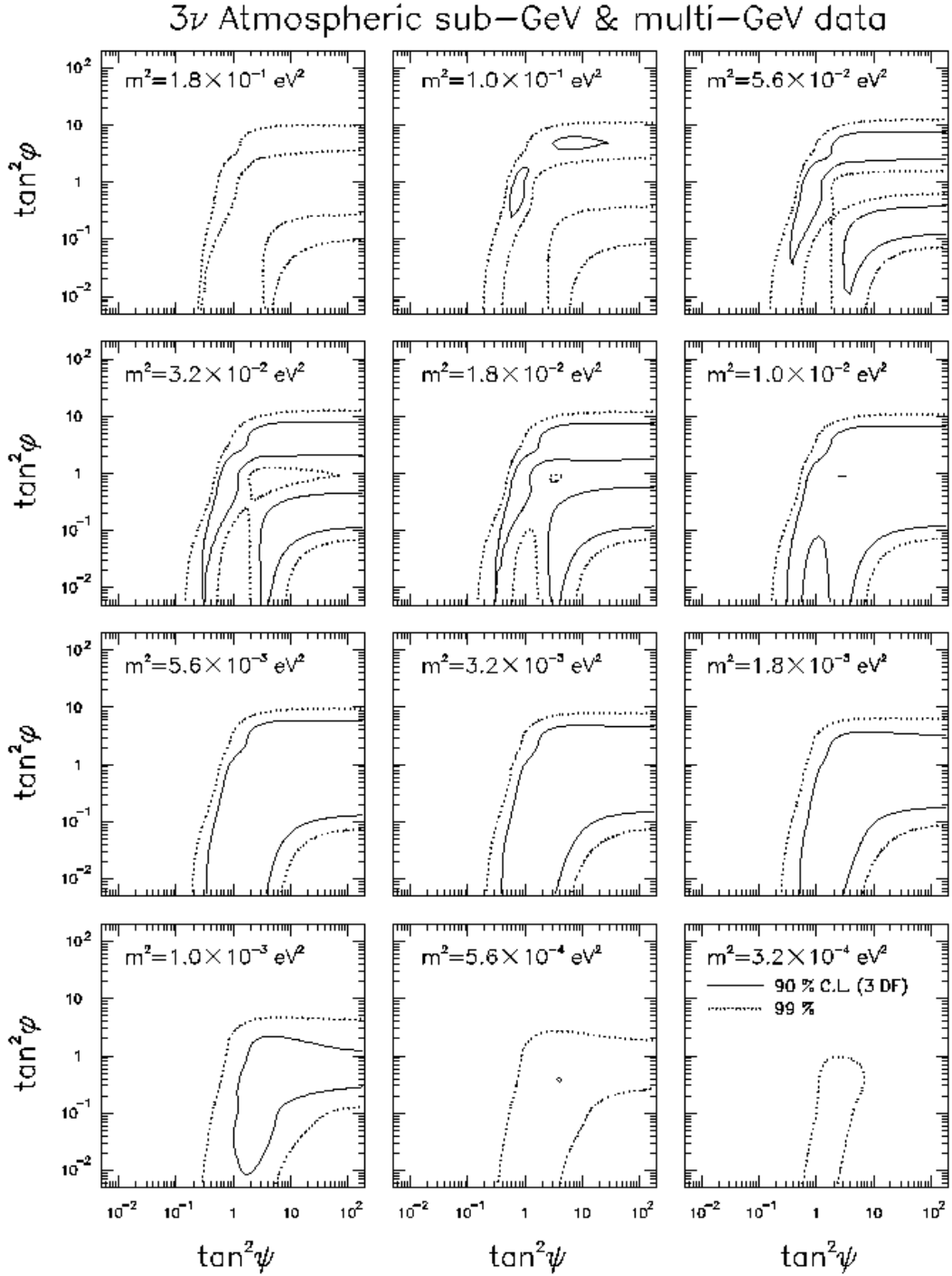

\caption{	Three-flavor analysis of all the atmospheric neutrino data
		(sub-GeV and binned multi-GeV combined) in the plane
		$(\tan^2\psi,\,\tan^2\phi)$, for 12 different values
		of $m^2$ ranging from $1.8\times10^{-1}$ to 
		$3.2\times10^{-4}$ eV$^2$. Scenario (a) of Fig.~1
		is assumed. The solid (dotted) curves represent sections
		of the region allowed at $90\%$ ($99\%$) C.L. for 
		$N_{\rm DF}=3$ at given $m^2$. The right side of each panel
		corresponds asymptotically to pure 
		$\nu_\mu\leftrightarrow\nu_e$ oscillations; the lower side 
		to pure $\nu_\mu\leftrightarrow\nu_\tau$ oscillations.
		Three-flavor oscillations interpolate smoothly between these
		two limits.}
\end{figure}
%...........................................................................08
\begin{figure}
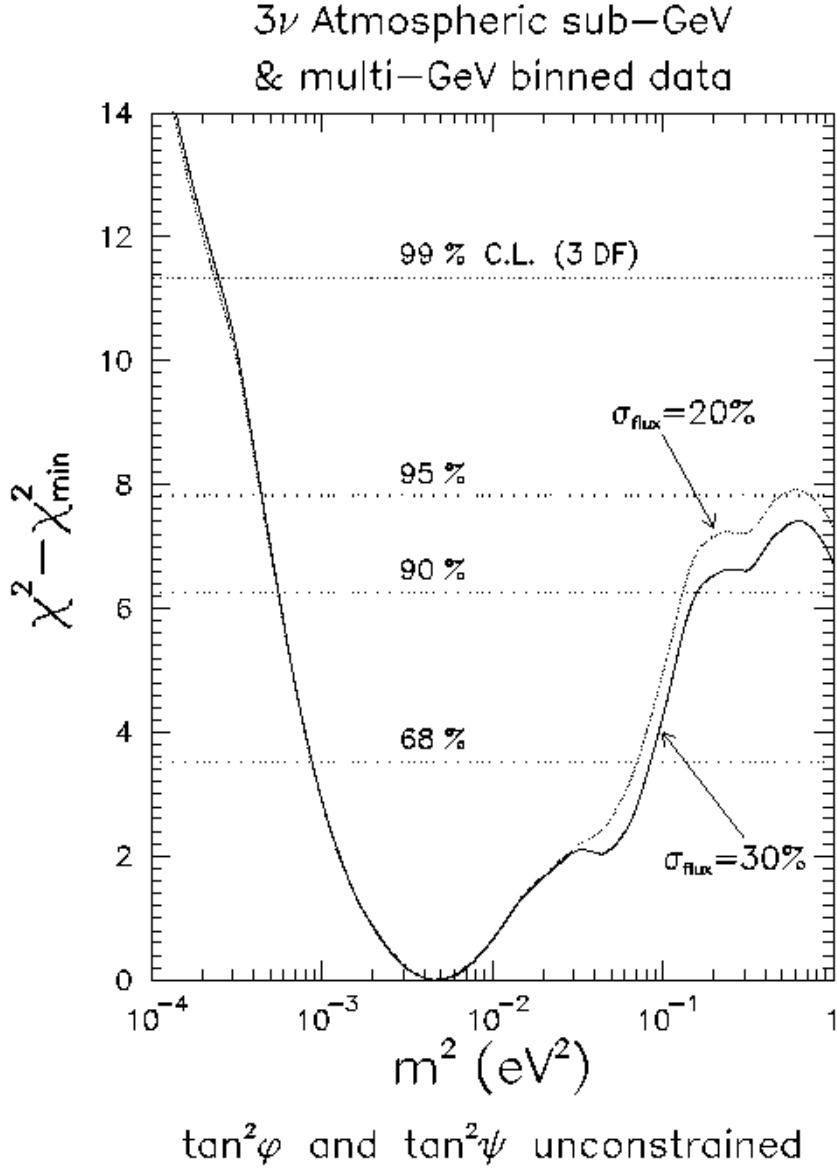

\caption{	Value of $\Delta\chi^2$ for all atmospheric neutrino data
		(sub-GeV and multi-GeV combined) as a function of $m^2$ 
		only. This figure embeds the information of Fig.~7
		{\em projected\/} onto the $m^2$ variable. At $68\%$
		C.L.\ ($N_{\rm DF}=3$) the value of $m^2$ is constrained 
		between $\sim 10^{-3}$ and $\sim 10^{-1}$ eV$^2$. For very 
		large $m^2$, the value of $\Delta\chi^2$ tends to $\sim 7$
		(not shown) and there are no upper limits on $m^2$ at
		$95\%$ C.L. The reduction of the $\nu$ flux error from
		$30\%$ to $20\%$ does not produce significant variations,
		as indicated by the thin, dotted curve.}
\end{figure}
%...........................................................................09
\begin{figure}
\caption{	As in Fig.~7, but in the scenario (b) of Fig.~1.}
\end{figure}
%...........................................................................10

\begin{figure}
\caption{	As in Fig.~8, but in the scenario (b) of Fig.~1.}
\end{figure}
%...........................................................................11
\begin{figure}
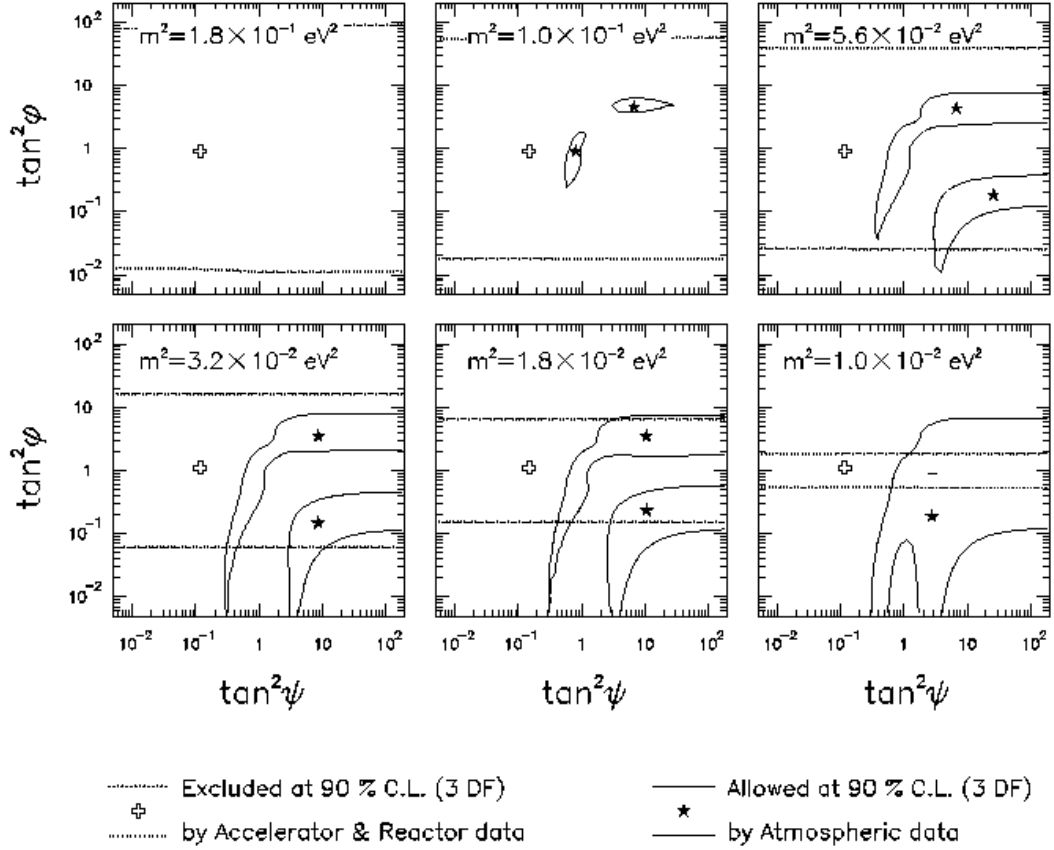

\caption{	Comparison between the regions allowed at $90\%$ C.L.\
		$(N_{\rm DF}=3)$ by the atmospheric neutrino data in 
		scenario (a) (solid contours), and the corresponding
		regions excluded by the established accelerator and reactor
		neutrino oscillation searches (horizontal, dotted contours). 
		Pure $\nu_\mu\leftrightarrow\nu_e$ atmospheric $\nu$ 
		oscillations (right side of each panel) are excluded by 
		accelerator and reactor data for 
		$m^2\protect\gtrsim 2\times 10^{-2}$ eV$^2$.
		There are no significant limits below $\sim10^{-2}$ eV$^2$
		from present accelerator and reactor searches.}
\end{figure}
%...........................................................................12
\begin{figure}
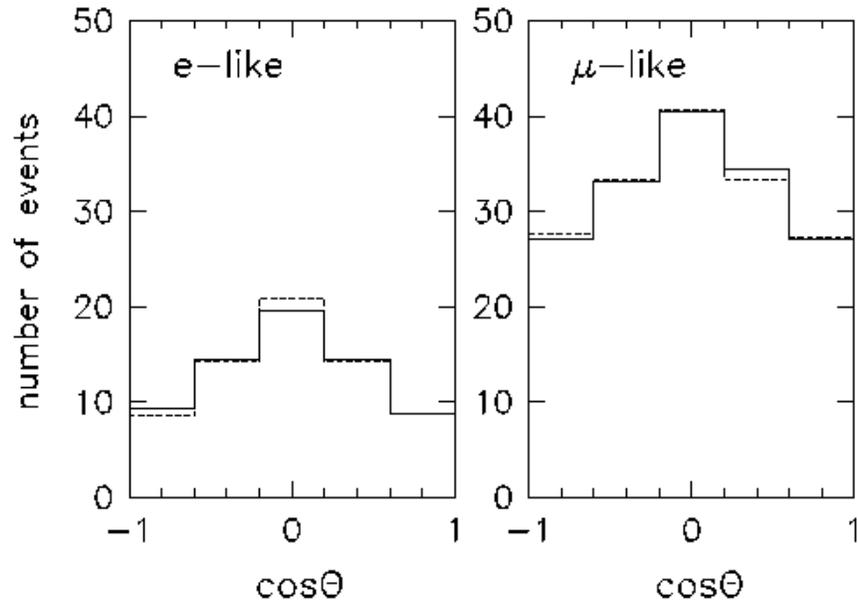

\caption{	Kamiokande  distribution of multi-GeV electrons and muons
		as a function of the zenith angle $\theta$, in absence of 
		neutrino oscillations. The agreement between the published 
		Kamiokande simulation (solid histogram) and our calculation 
		(dashed histogram) is very good. See Appendix~A for details.}
\end{figure}
%.............................................................................

%%%%%%%%%%%%%%%%%%%%%%%%%%%%%%%%%%%%%%%%%%%%%%%%%%%%%%%%%%%%%%%%%%%%%%%%%%%%%%
%\end{document}%       E N D      O F      R E V T E X       F I L E
%%%%%%%%%%%%%%%%%%%%%%%%%%%%%%%%%%%%%%%%%%%%%%%%%%%%%%%%%%%%%%%%%%%%%%%%%%%%%%

%%%%%%%%%%%%%%%%%%%%%%%%%%%%%%%%%%%%%%%%%%%%%%%%%%%%%%%%%%%%%%%%%%%%%%%%%%%%%%%
%%%%%%% 
%%%%%%%            THE FOLLOWING FOR AUTHOR USE ONLY.
%%%%%%%
%%%%%%%            INCLUSION OF BITMAPPED FIGURES WITH EPSFIG.STY.
%%%%%%%
%%%%%%%%%%%%%%%%%%%%%%%%%%%%%%%%%%%%%%%%%%%%%%%%%%%%%%%%%%%%%%%%%%%%%%%%%%%%%%%
%%%%%%%          P O S T S C R I P T       F I G U R E S 
%%%%%%%
%%%%%%%   memo:  1) add epsfig in the \documentstyle
%%%%%%%          2) and move this part befor \end{document} 
%%%%%%%		 3) remove previous figure captions
%%%%%%%          4) include the following \newcommand:
%%----------------------------------------------------------------------------
\newcommand{\InsertFigure}[2]{\newpage\begin{center}\mbox{%
\epsfig{bbllx=1.4truecm,bblly=1.3truecm,bburx=19.5truecm,bbury=26.5truecm,%
height=21.7truecm,figure=#1}}\end{center}\vspace*{-2.55truecm}%
\parbox[t]{\hsize}{\small\baselineskip=0.5truecm\hskip0.5truecm #2}}
%----------------------------------------------------------------------------
\newcommand{\InsertBitmap}[2]{\newpage\begin{center}\vspace*{-1.1cm}\mbox{%
\epsfig{bbllx=1.4truecm,bblly=1.3truecm,bburx=19.5truecm,bbury=26.5truecm,%
height=22.7truecm,figure=#1}}\end{center}\vspace*{-2.7truecm}%
\parbox[t]{\hsize}{\small\baselineskip=0.5truecm\hskip0.5truecm #2}}
%----------------------------------------------------------------------------
%%%%%%%%%%%%%%%%%%%%%%%%%%%%%%%%%%%%%%%%%%%%%%%%%%%%%%%%%%%%%%%%%%%%%%%%%%%%%%%
%..............................................................................
\InsertBitmap{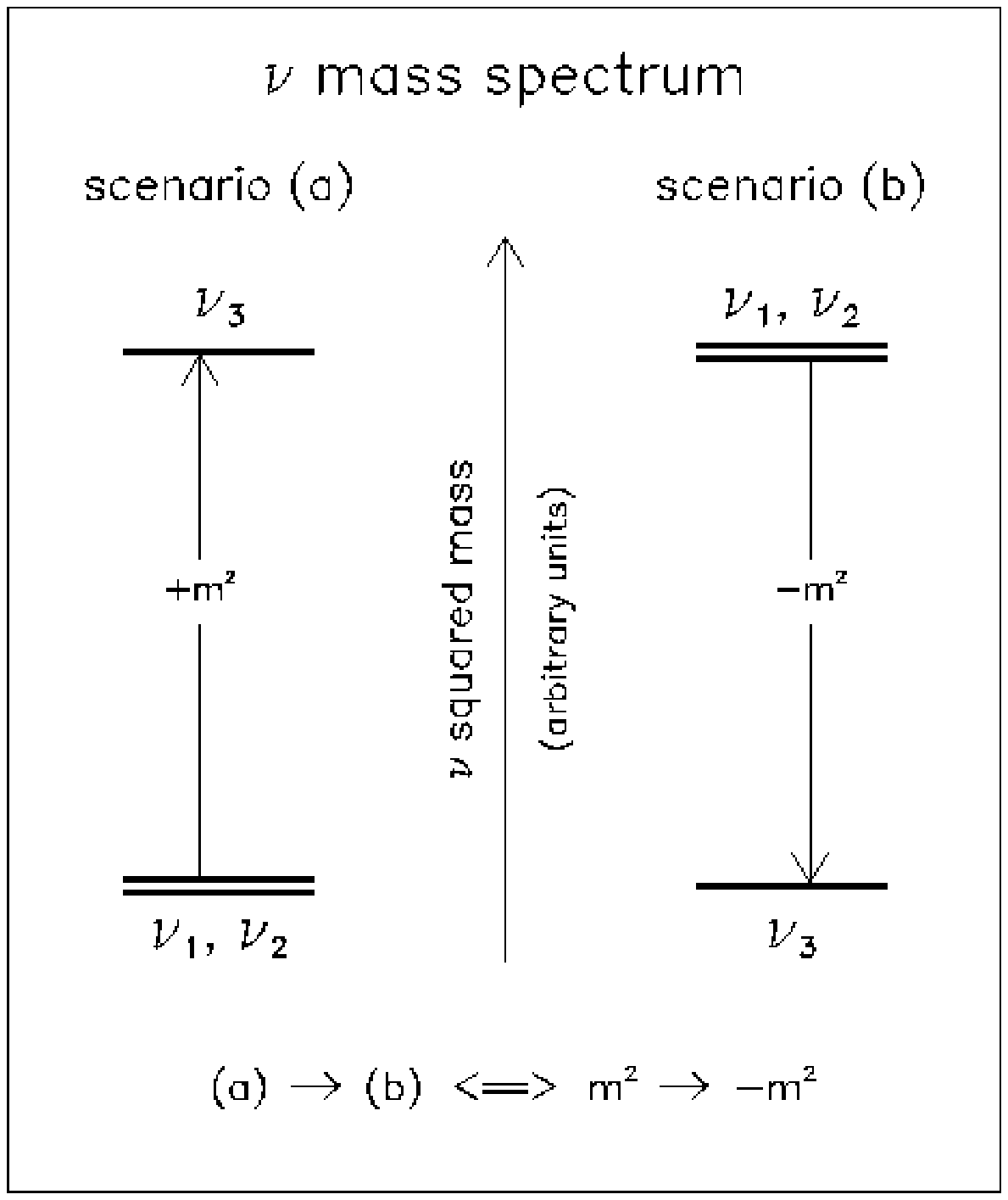}%
{FIG.~1. 	The neutrino mass spectrum adopted in this work. One 
		neutrino mass eigenstate $(\nu_3)$ is assumed to be 
		largely separated from the quasi-degenerate doublet 
		$(\nu_1,\,\nu_2)$ by a square mass difference
		$|m^2_3-m^2_{1,2}|\simeq m^2$.  The two possible
		scenarios (a) and (b) are physically different when
		the earth matter effects are included in the atmospheric 
		$\nu$ propagation. The two square mass spectra in (a) and (b) 
		are related by: 
		${\rm (a)}\to{\rm (b)}\Longleftrightarrow m^2\to-m^2$.}
%..............................................................................
\InsertBitmap{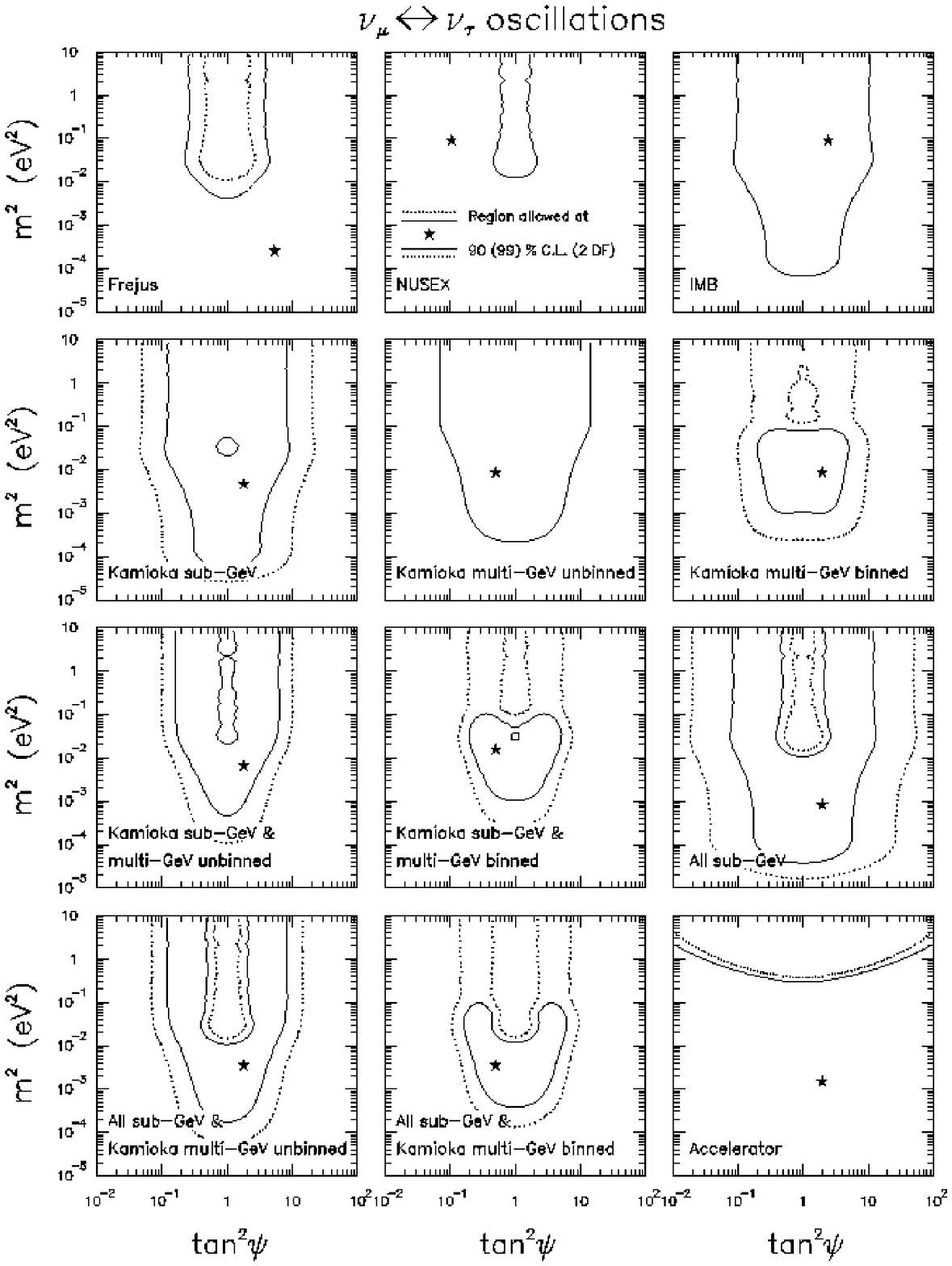}%
{FIG.~2. 	Analysis of separate and combined atmospheric neutrino
		data assuming pure $\nu_\mu\leftrightarrow\nu_\tau$
		oscillations ($\phi=0$ in our framework) in the mass-mixing
		plane $(\tan^2\psi,\,m^2)$.  Contours at $90\%$ and $99\%$
		C.L.\ for $N_{\rm DF}=2$ are shown. The allowed regions
		are marked by stars. The last panel shows the constraints 
		coming from the combination of established accelerator 
		data. Notice the symmetry of all the allowed regions with 
		respect to the axis $\psi=\pi/4$.}
%..............................................................................
\InsertBitmap{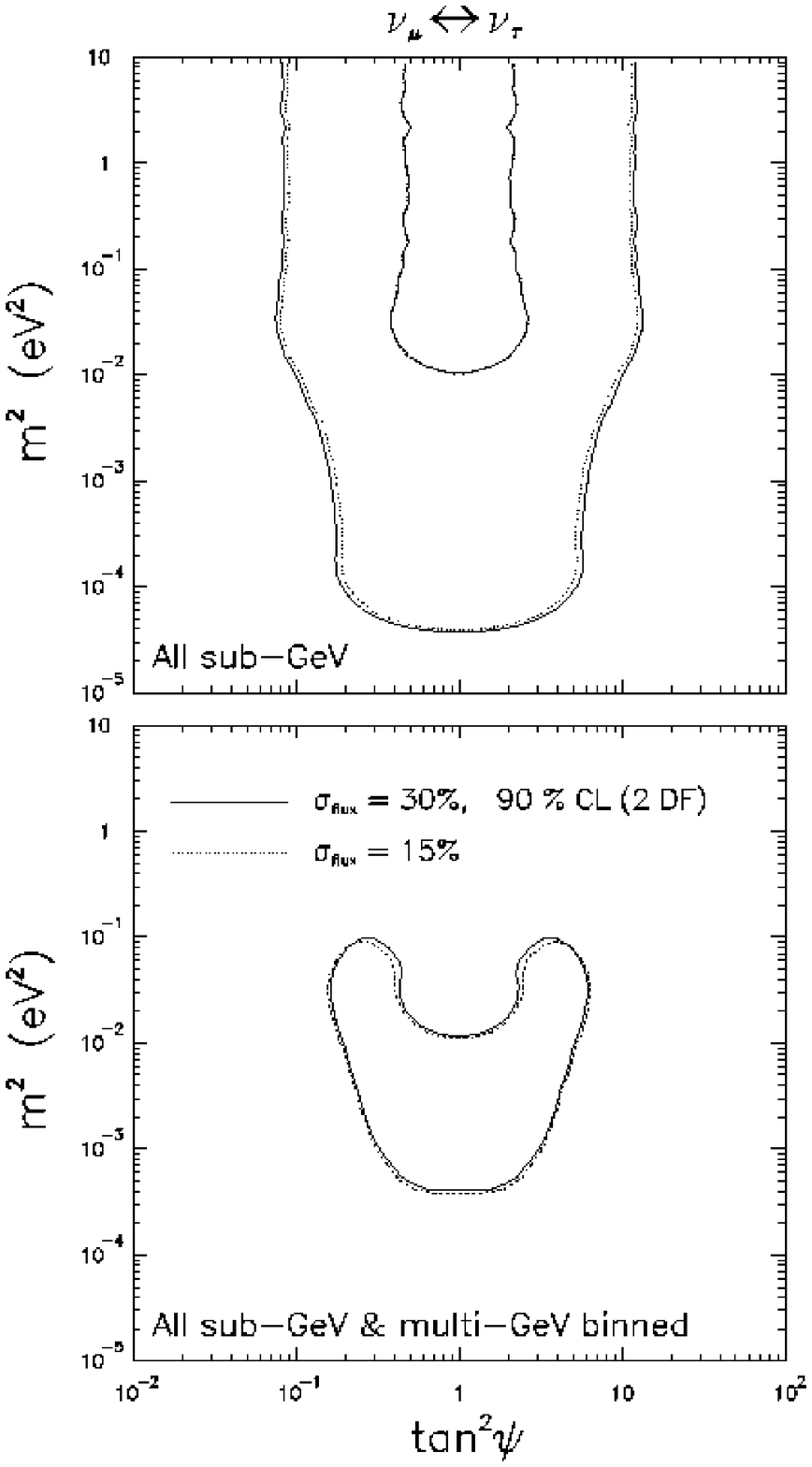}%
{FIG.~3.	Effects of the reduction of the neutrino flux uncertainty 
		from the default value of $\pm 30\%$ (solid lines) to 
		$\pm 15\%$ (dotted lines). The variations in the regions 
		allowed at $90\%$ C.L.\ by sub-GeV data (upper panel) 
		or sub-GeV+multi-GeV data (lower panel) are very small.}
%..............................................................................
\InsertBitmap{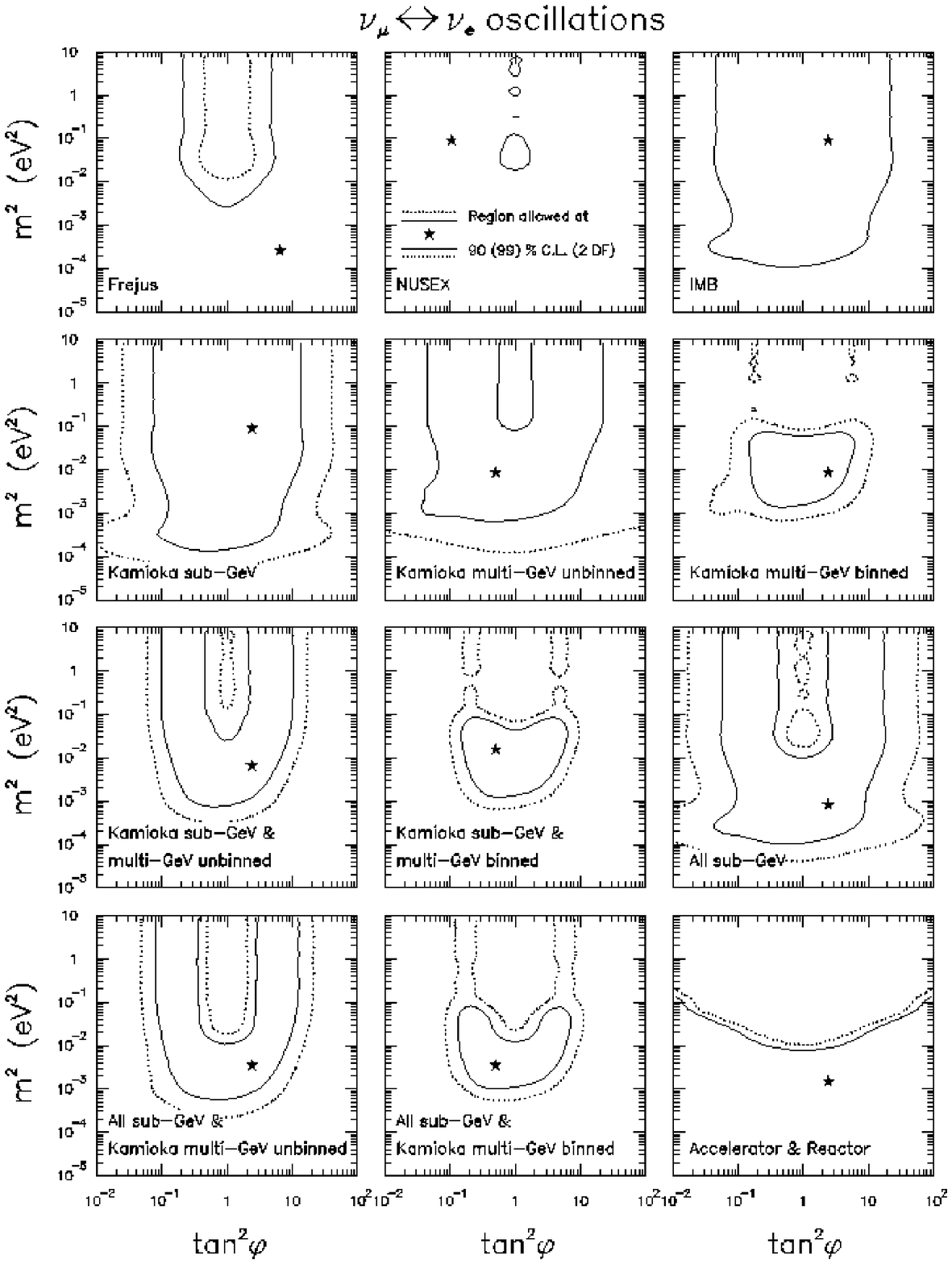}%
{FIG.~4.	Separate and combined analysis of atmospheric neutrino
		data assuming pure $\nu_\mu\leftrightarrow\nu_e$
		oscillations ($\psi=\pi/2$ in our framework), in the 
		mass-mixing plane $(\tan^2\phi,\,m^2)$. Contours at 
		$90\%$ (solid) and $99\%$ C.L.\ (dotted) for 
		$N_{\rm DF}=2$ are shown. The allowed regions are marked 
		by stars. The last panel shows the constraints coming 
		from the combination of established accelerator and 
		reactor data. The contours of the atmospheric $\nu$ 
		allowed regions are not symmetric with respect to the 
		axis $\phi=\pi/4$, due to matter oscillation effects.}
%..............................................................................
\InsertBitmap{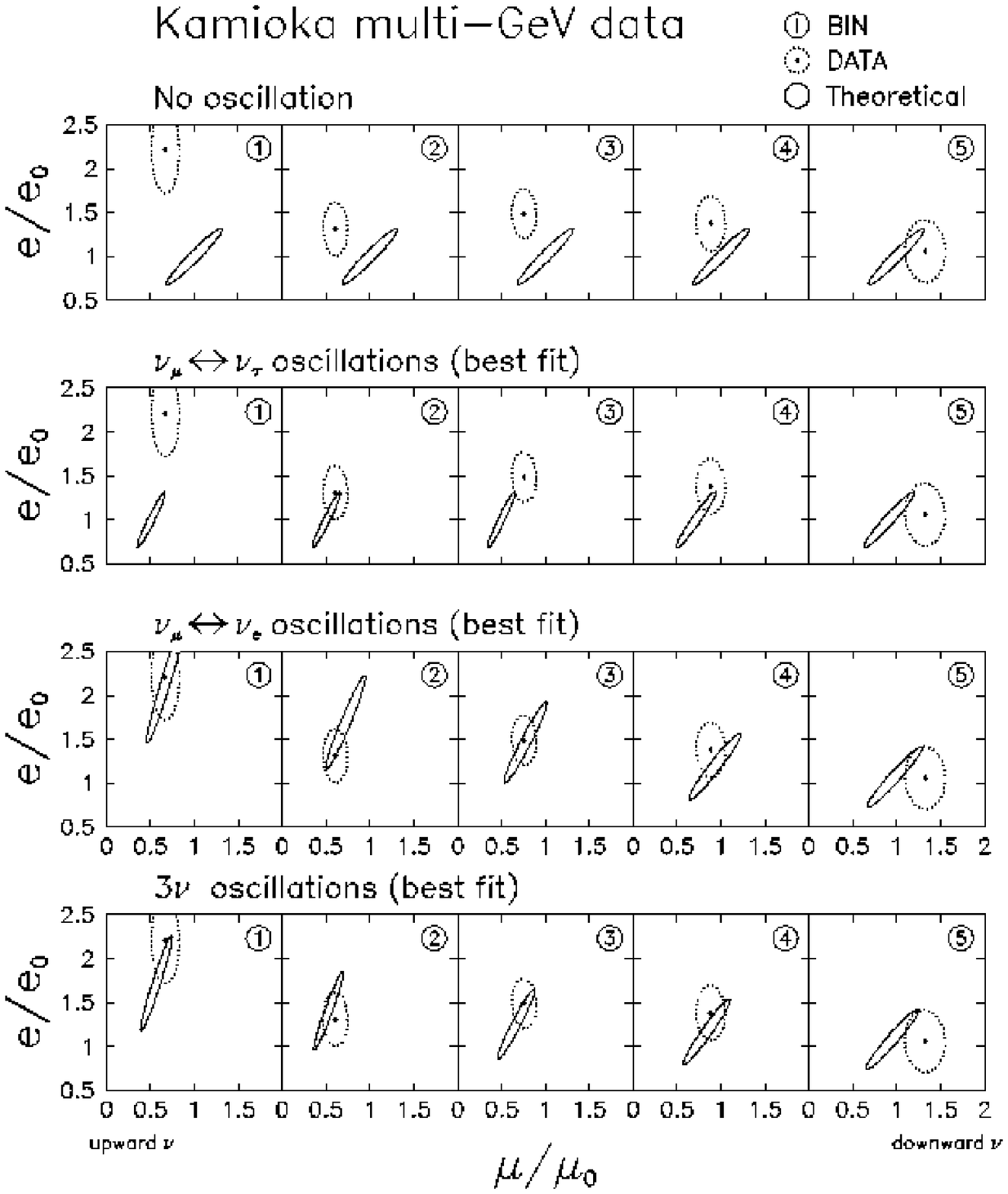}%
{FIG.~5.	Bin-by-bin analysis of multi-GeV Kamiokande data in
		the plane of the $\mu$ and $e$ lepton rates, normalized
		to their theoretical values without oscillations,
		$\mu_0$ and $e_0$. Solid ellipses: theoretical predictions 
		at $1\sigma$ level $(\Delta \chi^2=1)$. Dotted ellipses: 
		experimental data at $1\sigma$ level. Notice how the 
		theoretical ellipses change from the upper panel 
		(no oscillation) to the  three lower panels (best-fit 
		cases for two-flavor and three-flavor oscillations). The 
		fit is better in the $\nu_\mu\leftrightarrow\nu_e$ case 
		than in the $\nu_\mu\leftrightarrow\nu_\tau$ case. The
		overall fit improves slightly in the $3\nu$ oscillation case.}
%..............................................................................
\InsertBitmap{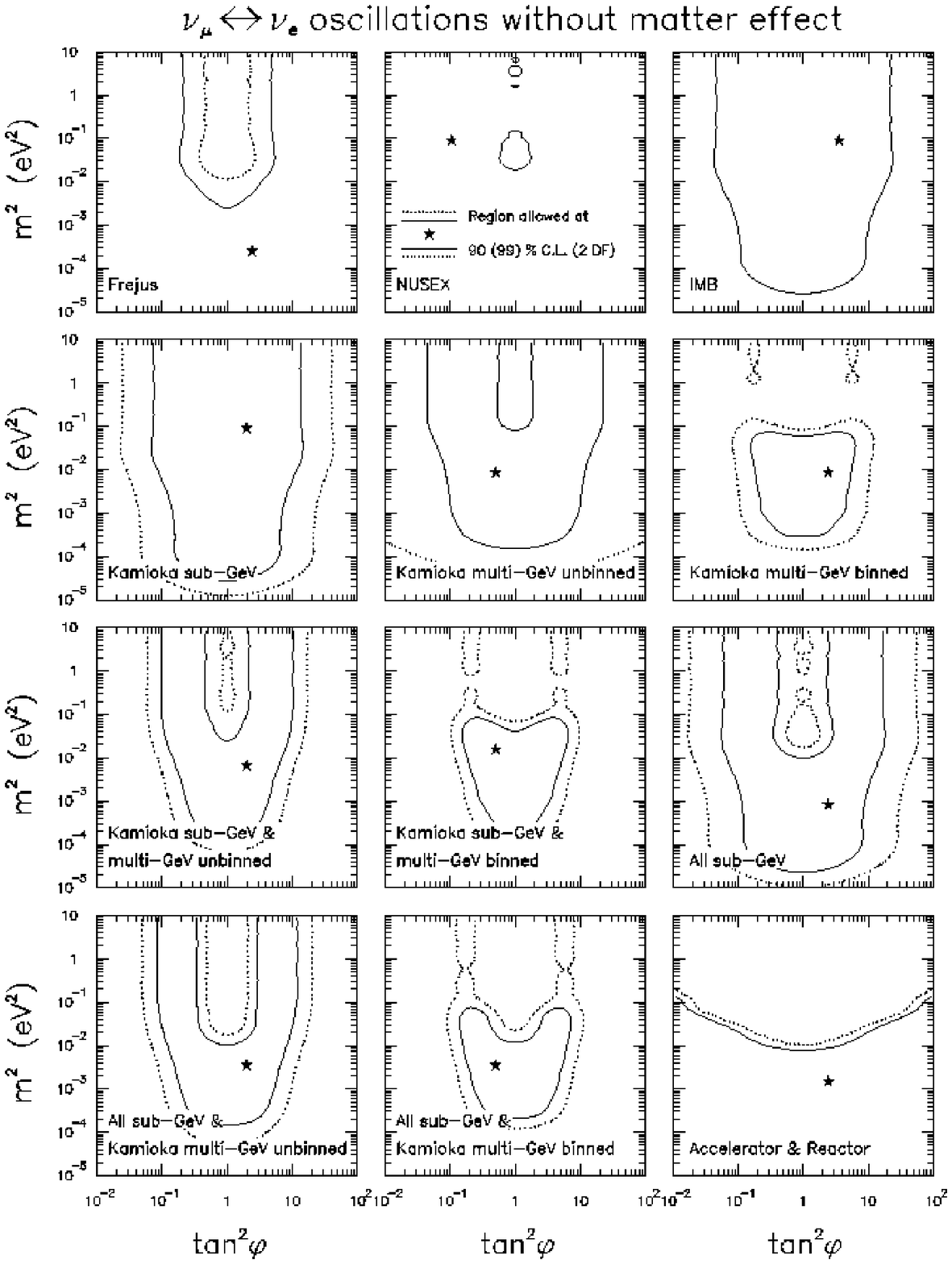}%
{FIG.~6.	As in Fig.~4, but excluding the earth matter effect
		(pure vacuum oscillations). All the contours are now 
		symmetric with respect to the axis $\phi=\pi/4$. The 
		regions allowed by the atmospheric neutrino data in the
		lower half of each panel are substantially different from 
		those reported in Fig.~4.}
%..............................................................................
\InsertBitmap{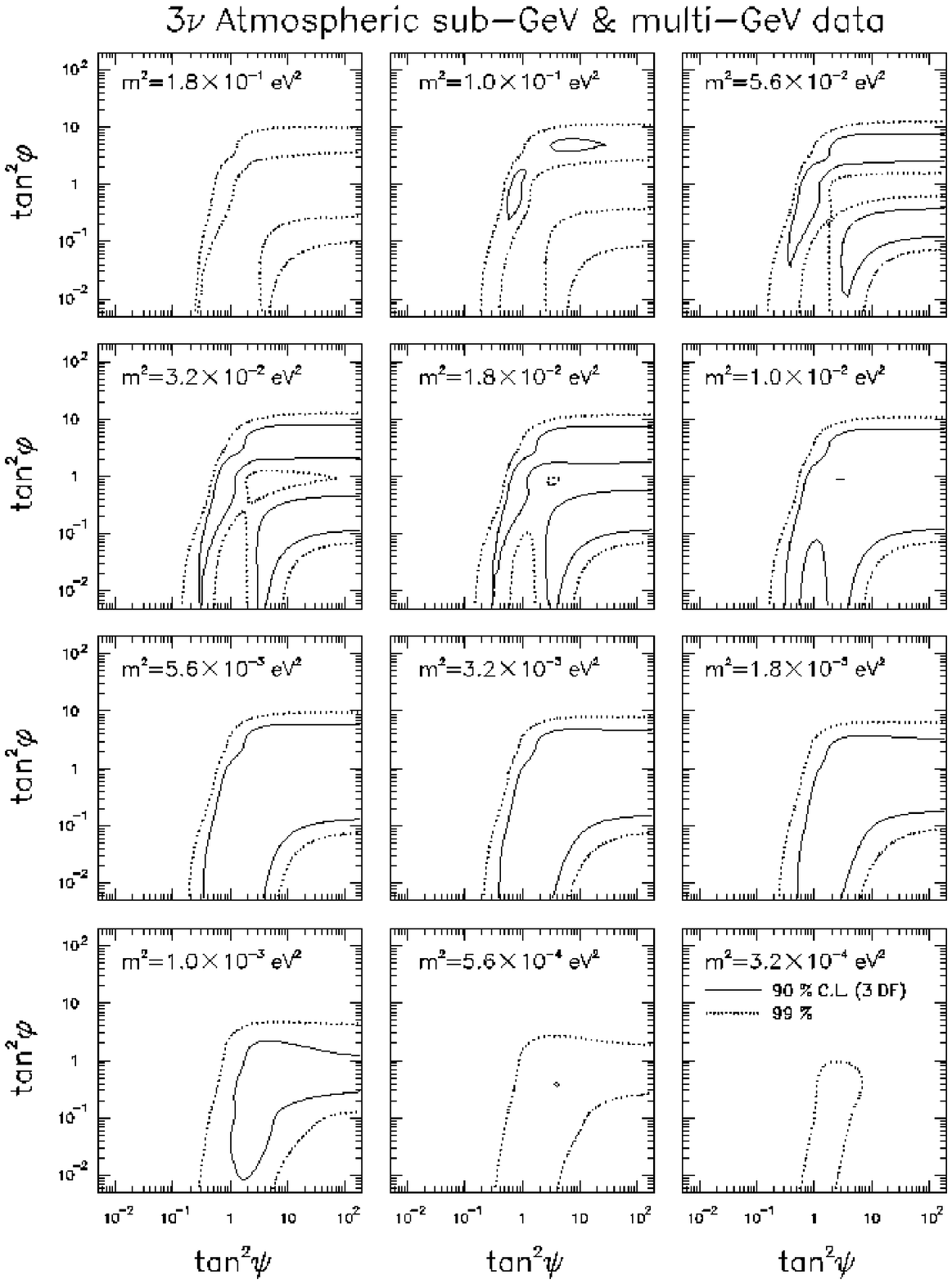}%
{FIG.~7.	Three-flavor analysis of all the atmospheric neutrino data
		(sub-GeV and binned multi-GeV combined) in the plane
		$(\tan^2\psi,\,\tan^2\phi)$, for 12 different values
		of $m^2$ ranging from $1.8\times10^{-1}$ to 
		$3.2\times10^{-4}$ eV$^2$. Scenario (a) of Fig.~1
		is assumed. The solid (dotted) curves represent sections
		of the region allowed at $90\%$ ($99\%$) C.L. for 
		$N_{\rm DF}=3$ at given $m^2$. The right side of each panel
		corresponds asymptotically to pure 
		$\nu_\mu\leftrightarrow\nu_e$ oscillations; the lower side 
		to pure $\nu_\mu\leftrightarrow\nu_\tau$ oscillations.
		Three-flavor oscillations interpolate smoothly between these
		two limits.}
%..............................................................................
\InsertBitmap{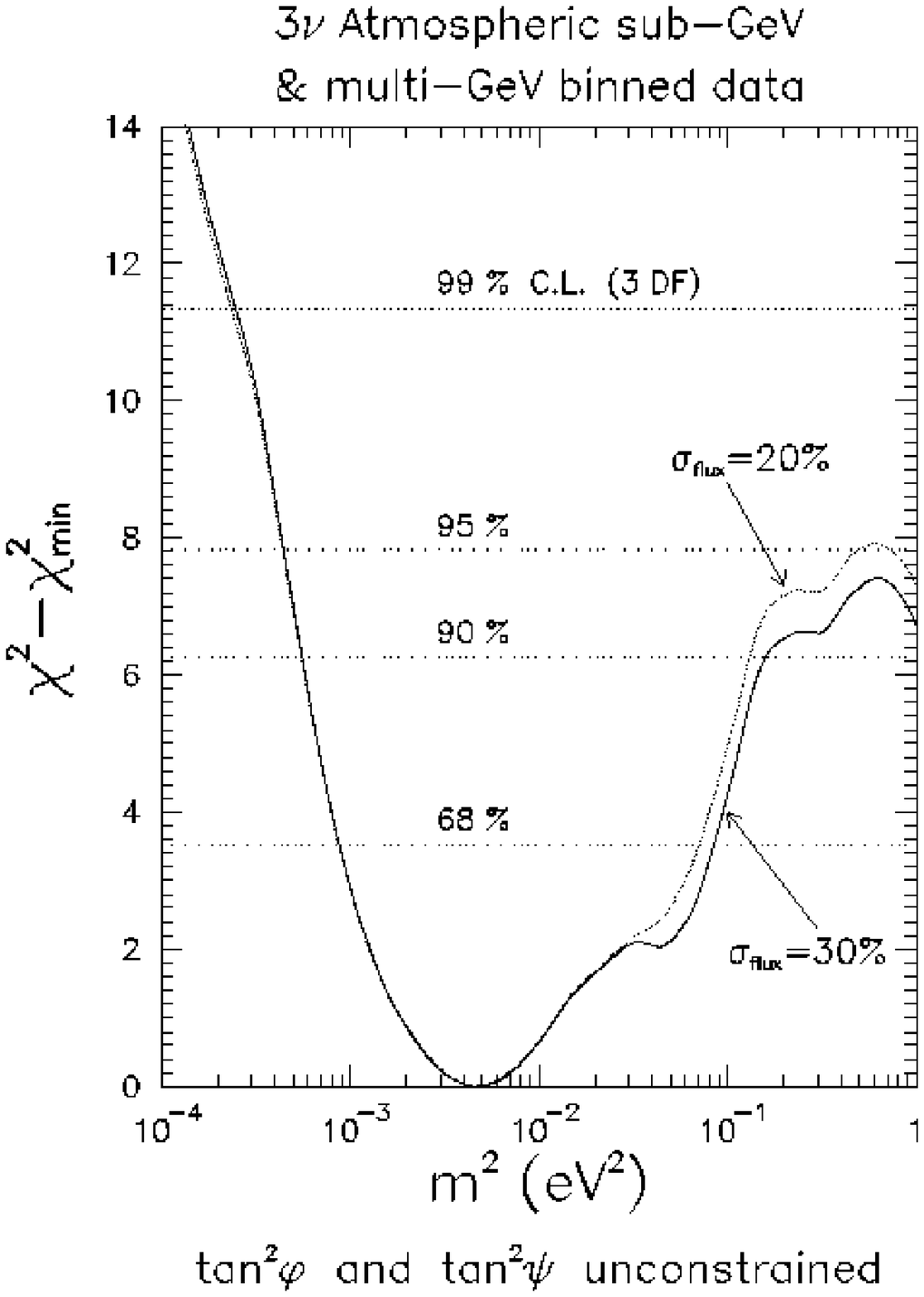}%
{FIG.~8.	Value of $\Delta\chi^2$ for all atmospheric neutrino data
		(sub-GeV and multi-GeV combined) as a function of $m^2$ 
		only. This figure embeds the information of Fig.~7
		{\em projected\/} onto the $m^2$ variable. At $68\%$
		C.L.\ ($N_{\rm DF}=3$) the value of $m^2$ is constrained 
		between $\sim 10^{-3}$ and $\sim 10^{-1}$ eV$^2$. For very 
		large $m^2$, the value of $\Delta\chi^2$ tends to $\sim 7$
		(not shown) and there are no upper limits on $m^2$ at
		$95\%$ C.L. The reduction of the $\nu$ flux error from
		$30\%$ to $20\%$ does not produce significant variations,
		as indicated by the thin, dotted curve.}
%..............................................................................
\InsertBitmap{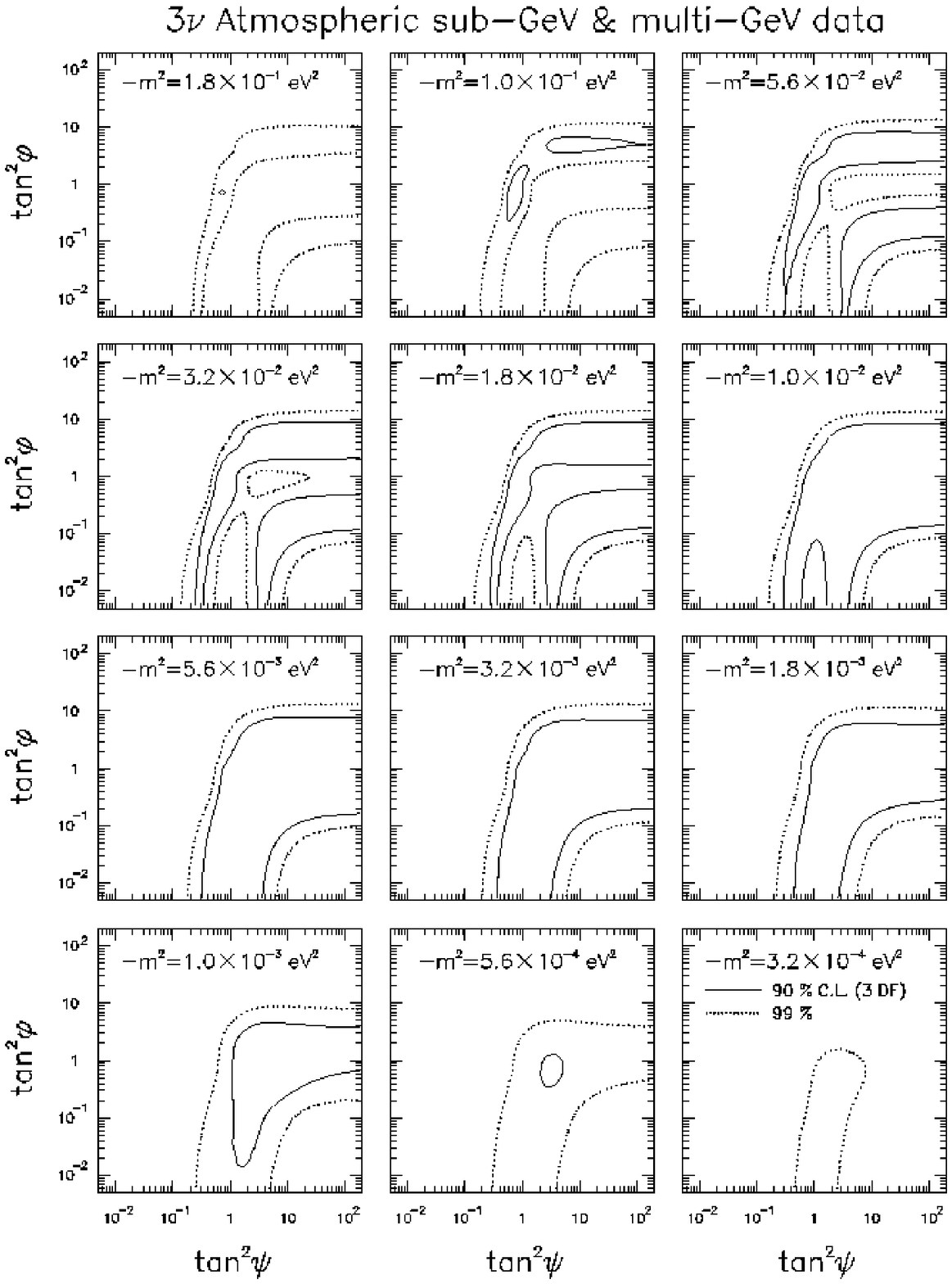}%
{\hfil FIG.~9.	As in Fig.~7, but in the scenario (b) of Fig.~1.\hfil}
%..............................................................................
\InsertBitmap{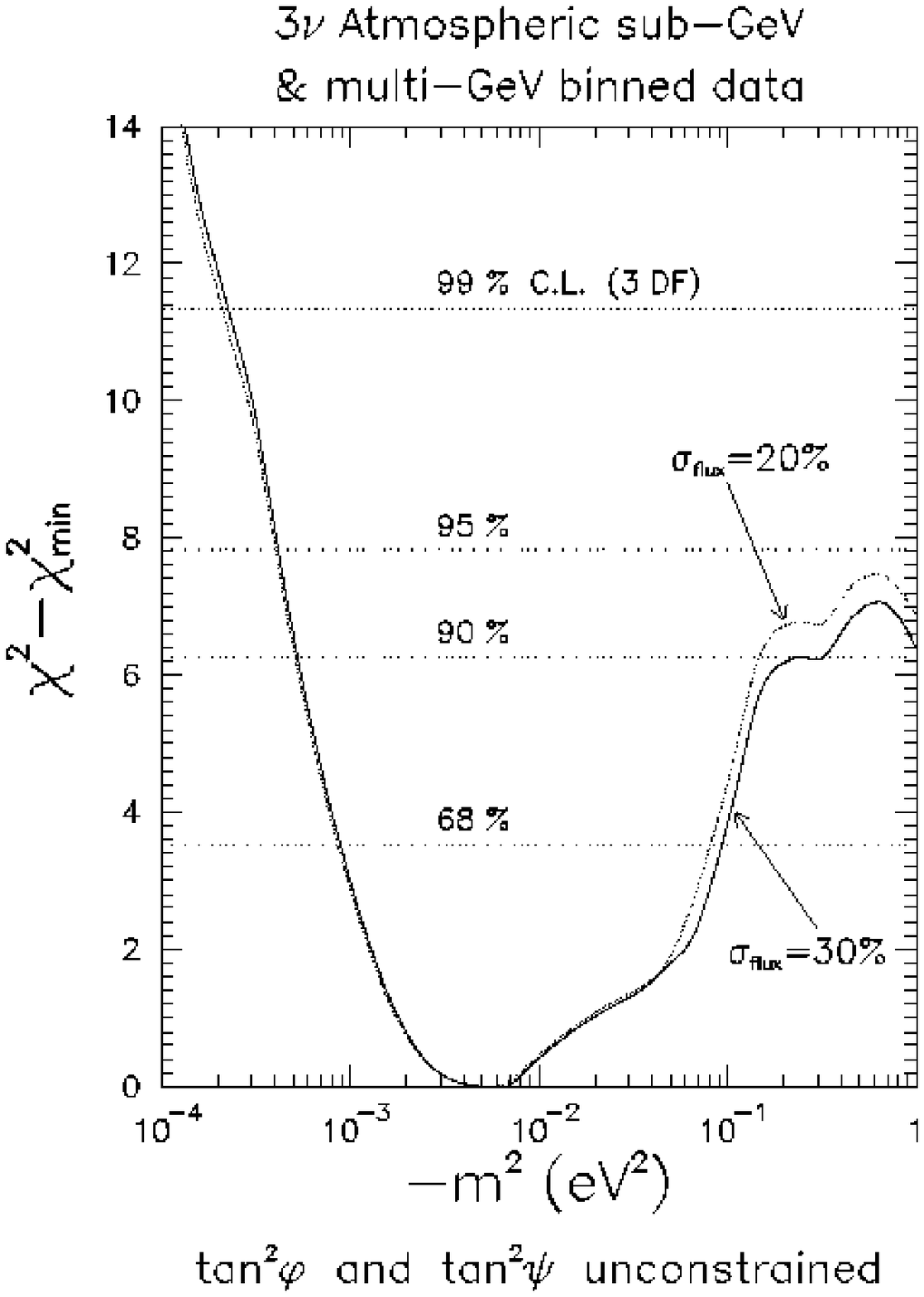}%
{\hfil FIG.~10.	As in Fig.~8, but in the scenario (b) of Fig.~1.\hfil}
%..............................................................................
\InsertBitmap{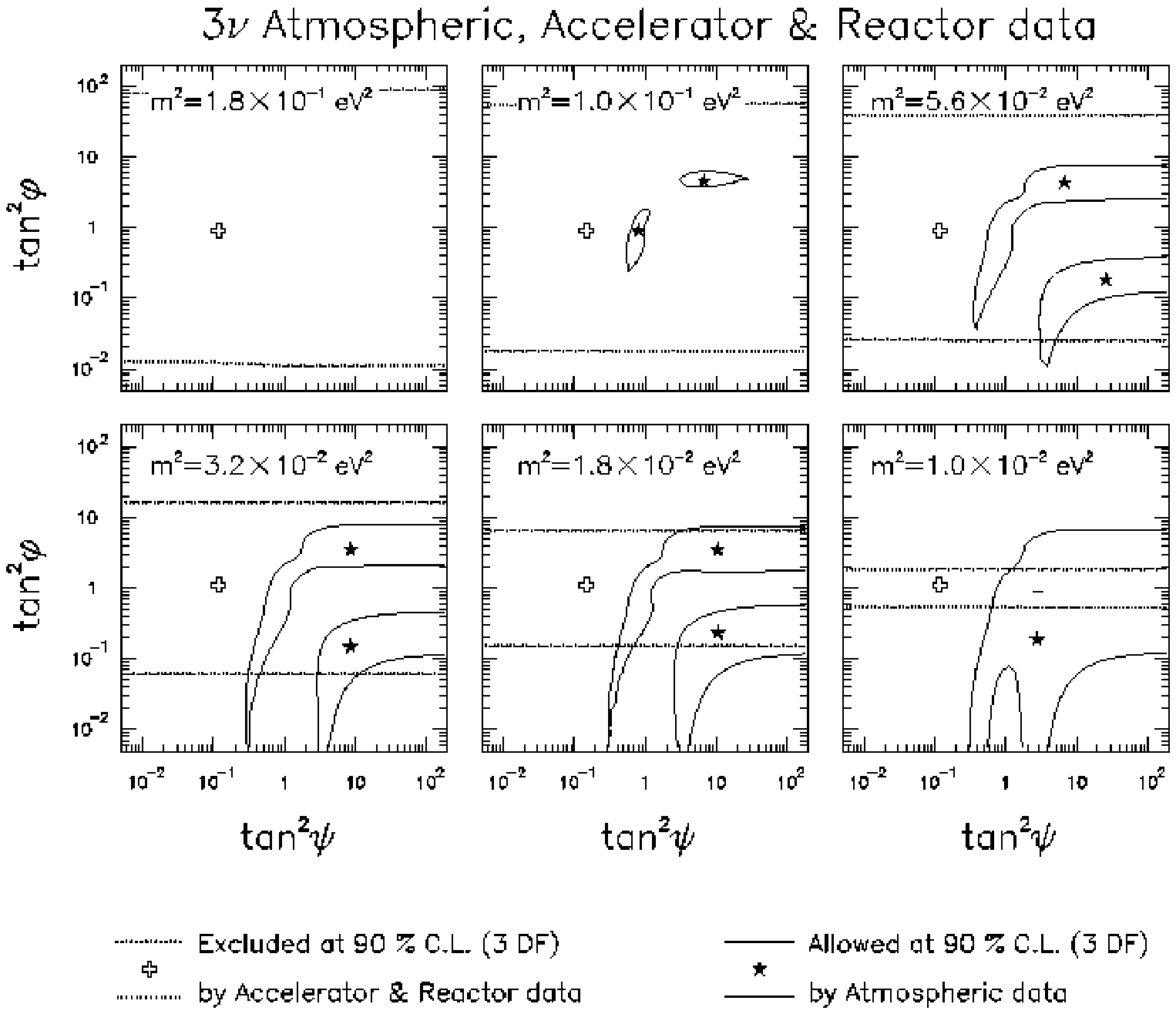}%
{FIG.~11.	Comparison between the regions allowed at $90\%$ C.L.\
		$(N_{\rm DF}=3)$ by the atmospheric neutrino data in 
		scenario (a) (solid contours), and the corresponding
		regions excluded by the established accelerator and reactor
		neutrino oscillation searches (horizontal, dotted contours). 
		Pure $\nu_\mu\leftrightarrow\nu_e$ atmospheric $\nu$ 
		oscillations (right side of each panel) are excluded by 
		accelerator and reactor data for 
		$m^2\protect\gtrsim 2\times 10^{-2}$ eV$^2$.
		There are no significant limits below $\sim10^{-2}$ eV$^2$
		from present accelerator and reactor searches.}
%..............................................................................
\InsertBitmap{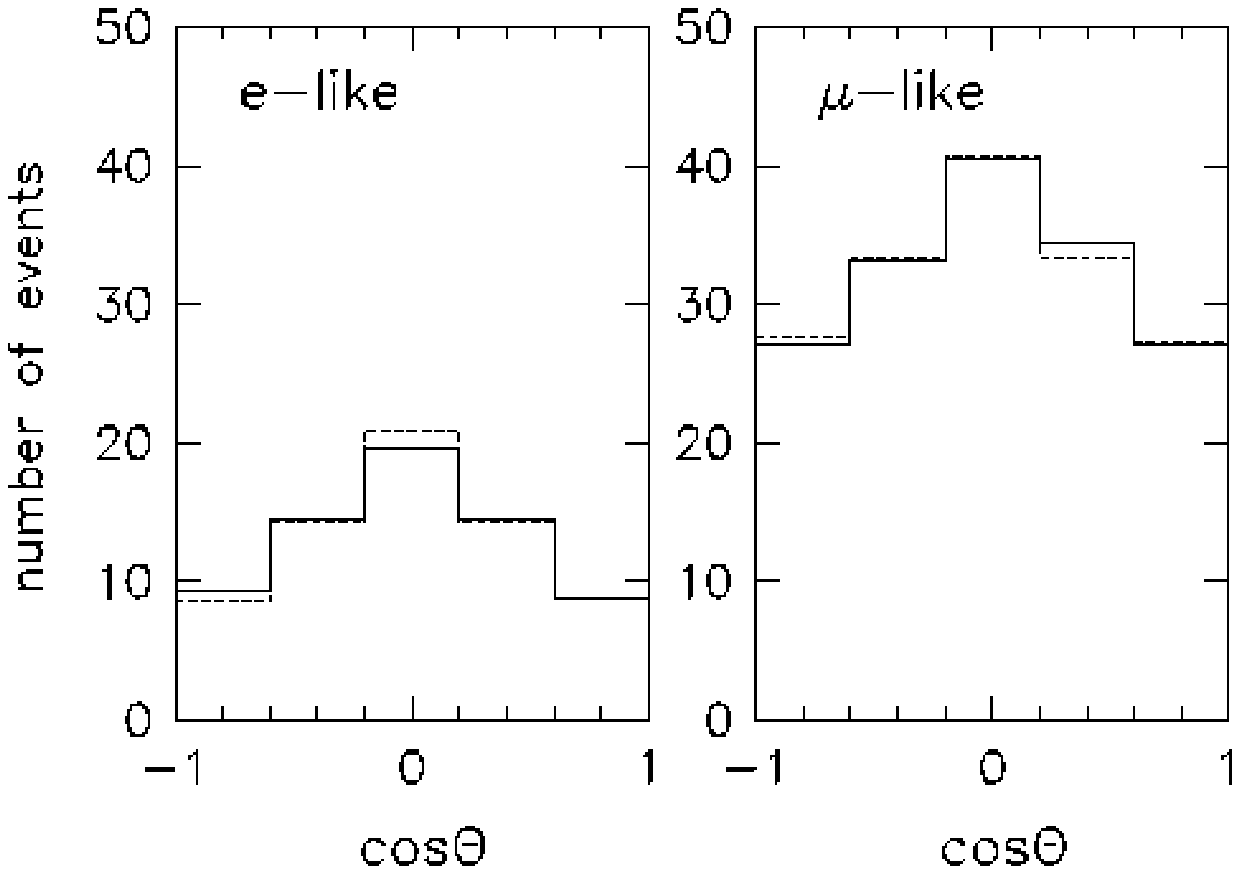}%
{FIG.~12.	Kamiokande  distribution of multi-GeV electrons and muons
		as a function of the zenith angle $\theta$, in absence of 
		neutrino oscillations. The agreement between the published 
		Kamiokande simulation (solid histogram) and our calculation 
		(dashed histogram) is very good. See Appendix~A for details.}
%%%%%%%%%%%%%%%%%%%%%%%%%%%%%%%%%%%%%%%%%%%%%%%%%%%%%%%%%%%%%%%%%%%%%%%%%%%
\end{document}